\documentclass[11pt]{article}

\usepackage{arxiv}

\usepackage{amsmath,amsfonts,latexsym,fullpage,graphicx,vmargin, mathrsfs}
\usepackage{pstricks}
\usepackage{comment}
\usepackage[inline]{enumitem}
\usepackage{cleveref}
\usepackage{float}

\usepackage[ruled]{algorithm2e}
\usepackage{algpseudocode}

\RequirePackage[OT1]{fontenc}

\usepackage{amsmath}
\usepackage{amsfonts}
\usepackage{caption}
\usepackage{subcaption}
\usepackage{multirow}
\usepackage{tikz}
\usetikzlibrary{graphs}
\usetikzlibrary{cd}
\usepackage{bm}
\usepackage{url}
\usepackage{placeins}

\DeclareMathOperator*{\ch}{C}
\DeclareMathOperator*{\im}{Im}
\DeclareMathOperator*{\h}{H}
\DeclareMathOperator*{\B}{B}
\DeclareMathOperator*{\Z}{Z}

\RequirePackage{natbib}

\newcommand{\virgolette}[1]{``#1''}

\usepackage{scalerel}

\newcommand{\R}{\mathbb{R}}

\usepackage[acronym]{glossaries}
\makeglossaries
\newacronym{aaa}{AAA}{Abdominal Aortic Aneurysm}
\newacronym{raaa}{RAAA}{Ruptured Abdominal Aortic Aneurysm}
\newacronym{ilt}{ILT}{intraluminal thrombus}
\newacronym{osi}{OSI}{Oscillatory Stress Index}
\newacronym{cta}{CTA}{Computed Tomography Angiography}
\newacronym{vmtk}{VMTK}{Vascular Modelling Toolkit}
\newacronym{mds}{MDS}{Multidimensional Scaling}

\usepackage[normalem]{ulem}

\firstpageno{1}

\begin{document}

\title{Persistence diagrams for exploring the shape variability of abdominal aortic aneurysms}

\author{Dario Arnaldo Domanin\thanks{Moxoff,Milano}, 
\and Matteo Pegoraro\thanks{Department of Mathematical Sciences, Aalborg University}, \and Santi Trimarchi\thanks{Department of Clinical Sciences and Community Health, Università degli Studi di Milano} \thanks{Dipartimento Area Cardio-Toraco Vascolare, S.C. Chirurgia Vascolare, Fondazione IRCCS Cà Granda Ospedale Maggiore Policlinico}, 
\and Maurizio Domanin\footnotemark[3] \footnotemark[4],
\and Piercesare Secchi\thanks{MOX -- Department of Mathematics, Politecnico di Milano}}

%
\maketitle

\abstract{
Abdominal Aortic Aneurysm consists of a permanent dilation in the abodminal portion of the aorta and, along with its associated pathologies like calcifications and intraluminal thrombi, is one of the most important pathologies of the circulatory system. The shape of the aorta is among the primary drivers for these health issues, with particular reference to all the characteristics which affects the hemodynamics. Starting from the computed tomography angiography of a patient, we propose to summarize such information using tools derived from Topological Data Analysis, obtaining persistence diagrams which describe the irregularities of the lumen of the aorta. We showcase the effectiveness of such shape-related descriptors with a series of supervised and unsupervised case studies.
}

\keywords{
topological data analysis; abdominal aortic aneurysm; persistent homology.}

\maketitle


\section{Introduction}\label{sec:intro}

One of the main drivers of contemporary medical research is the idea of \emph{personalized medicine}, which aims at providing better targeted treatments to improve therapy outcomes. 
The achievement of such a challenging task relies primarily on two factors: the constantly increasing amount of data made available by modern data collecting pipelines, like medical imaging, and the joint efforts of clinicians and statisticians, trying to unpack the valuable information contained in the provided data.

The statistical understanding of any phenomenon is always limited by the interpretable methods and algorithms that the analyst can resort to. As a consequence, there is an increasing need of new and original statistical methods aimed at the analysis of different and heterogeneous kinds of complex data, validated and interpreted by the expert clinician. 
In many medical situations, especially those related to imaging data, many complexities arise because of the difficulty in decoupling interesting variability between statistical units from that which the clinician considers as ancillary. This problem is at the root of \emph{object oriented data analysis} \cite{marron2021object}, whose foundational principles focus on data representation, by embedding the atoms of the statistical analysis in a suitable mathematical space where their variability of interest can be captured and explored.

In this work we consider the problem of analysing and characterising \emph{abdominal aortic aneurysms} (\acrshort{aaa}s)
eliciting the information contained in the shape of blood vessels with tools from algebraic topology.

\begin{figure} 

\begin{subfigure}{0.5\textwidth} 
    \centering
    \includegraphics[width=\textwidth]{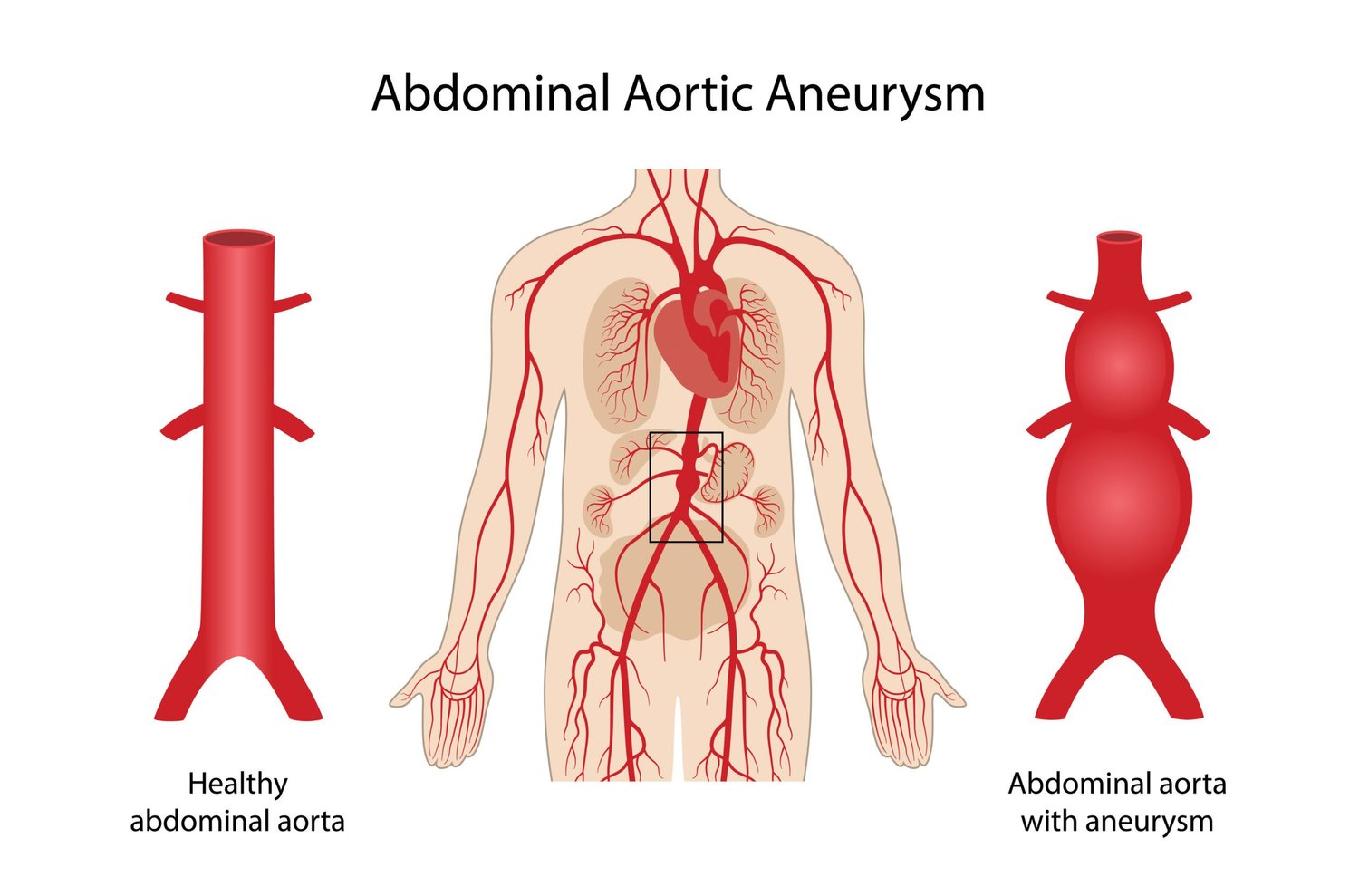}
    \caption{Schematic view of the circulatory system with highlighted the abdominal portion of the aorta, with both an healthy and aneurysmatic example. Image from \protect\url{https://www.vascularcures.org/abdominal-aortic-aneurysms}.}
    \label{fig:AAA}
\end{subfigure}
\begin{subfigure}{0.46\textwidth} 
    \centering
    \includegraphics[width=\textwidth]{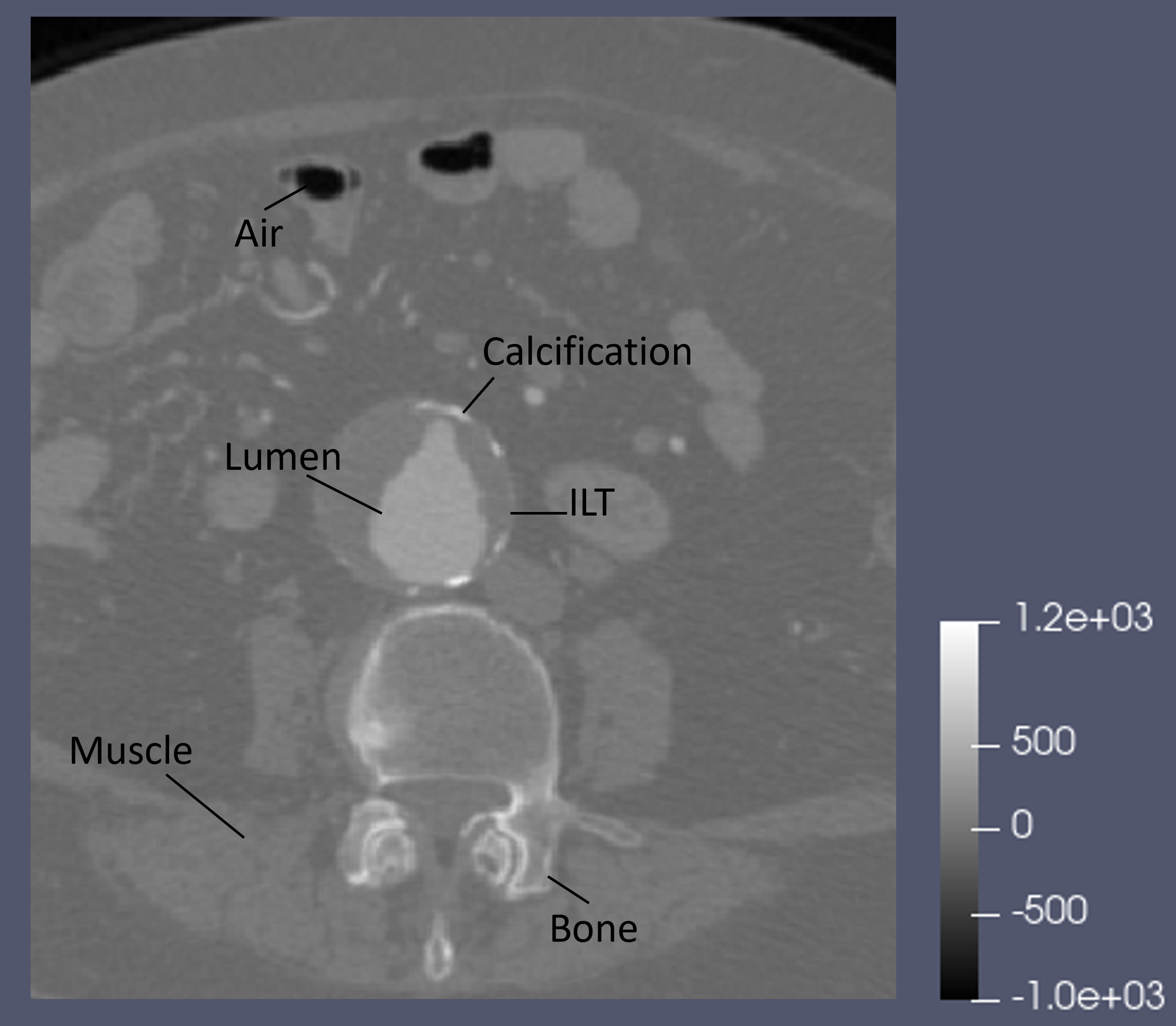}
    \caption{Coronal slice of a CTA scan. Different shades of grey differentiate between different organs and tissues. This allows both the clinicians to see the entire anatomy of the patient and the algorithm to execute the segmentation process.}
    \label{fig:CTA}
\end{subfigure}

\caption{Schematic representation of an \acrlong{aaa} and a CTA scan.}
\end{figure}

\acrlong{aaa} consists of a permanent dilation of at least 1.5 times the expected diameter of the abdominal portion of the aorta (see \Cref{fig:AAA}), the biggest artery of the human body \cite{1_intro_01}, characterized by chronic inflammation processes in the inner layers with degradation or redistribution of elastin and collagen \cite{1_intro_05} under the haemodynamical stress \cite{1_intro_06}.
Such dilatation can evolve asymptomatically towards the progressive enlargement of the vessel up to its rupture which is frequently fatal, making it one of the most prominent pathologies of the circulatory system.

Most of the literature dealing with the analysis of \acrshort{aaa}s focuses on the growth rate of small AAAs \cite{parr2011comparison, kauffmann2012measurements}, in order to prevent their rupture and identify the necessity of surgery, leveraging on shape-related or biomechanical numerical features. See, for instance, \cite{1_intro_08, 1_intro_09, 1_intro_10, zhu2020intraluminal, lindquist2021geometric}. Differences between these works appear in the statistical methodology employed for prediction, but, mostly, in the kind of features collected: \acrshort{aaa} and intraluminar thrombus' diameter, volume, axes are considered in most studies; \cite{1_intro_10}, \cite{lindquist2021geometric} and others include also biomechanical indices - like peak wall stress and peak wall rupture indices - and clinical variables, showing improved predictions.

The aim of our work is different and is more in line with \cite{kyriakou2019methodology}; indeed, we want to estabilish a new mathematical representation of the geometric complexity of the aorta and of the surrounding blood vessels. A deep dive on the heuristic power of this new representation is the main aim of this paper; we leave to future work the assessment of its predictive power in terms of growth rate of small \acrshort{aaa}s.   
Nevertheless, most of the information conveyed by morphological and shape related variables is also contained in our novel representation, which, however, has the extra advantage of summarizing additional information which is harder to convey in terms of numerical variables; like, for instance, a quantification of the calcifications along the blood vessels.

To obtain such representation we resort to algebraic topology \cite{munkres2018elements} and, in particular, to \emph{persistent homology} \cite{edelsbrunner2022computational}, one of the most diffused frameworks in Topological Data Analysis (TDA). 
By introducing a carefully studied and application-driven filtering function, we are able to synthesize the Computed Tomography Angiographies (\acrshort{cta}s) of patients - a particular kind of medical images, see \Cref{fig:CTA} - capturing the statistically sufficient features of their \emph{shapes}. Moving from a preliminary segmentation pipeline, already well established and whose goal is to turn CTA images into 3D meshes, we transform them into mathematical objects called \emph{persistence diagrams}. 
Thus, the statistical analysis is moved from the initial CTA images to their representations embedded in the space of persistence diagrams. To demonstrate the effectiveness of this representation we propose a number of analyses characterized by the simplicity of the pipeline, but heavily relying on the power of the chosen representation.

The approach pursued in this work paves the way for more refined analyses - for instance on the growth rate of small AAAs - grounded on the same topological representation and aimed at providing clinicians with valuable insights and checkpoints to complement their current analyses. We also foresee applications of analogous filtering functions to other contexts needing similar shape-dependent characterizations, such as cerebral aneurysm and 
the development of the atheromatous plaque in arterial vessels.

\subsection{Outline}

In \Cref{sec:AAA_intro} we give a brief overview of Abdonimal Aortics Aneurysms and their related pathologies. Then, in \Cref{sec:data}, we concisely describe the segmentation pipeline to obtain the meshes from the CTA scans. \Cref{sec:PH} is devoted to presenting the (normalized) radial filtration function and the associated persistence diagrams. \Cref{sec:read_PDs} is entirely focused on reading patient-related information from persistence diagrams. In \Cref{sec:analyses} we propose a number of data analysis situations which are easily tackled with the use of persistence diagrams. All the packages and softwares we employ for visualization and analyses are reported in \Cref{sec:softw}. Finally, \Cref{sec:discussion} ends the manuscript with a discussion and a conclusion. The Appendix contains some further technical details.

\section{Abdominal Aortic Aneurysm}\label{sec:AAA_intro}

\acrlong{aaa} is the most common aortic pathology. The incidence of \acrlong{raaa} (\acrshort{raaa}) is up to 17.5 per 100,000 person/year in Western countries \cite{1_intro_03}. It is mainly located between the renal and the iliac arteries, which can also present dilatations on their owns called \emph{iliac aneurysms}. The healthy segment of the abdominal aorta above the dilatation is defined as \textit{aortic neck}.

Two main features characterize the aneurysm: \textit{\acrlong{ilt}} (\acrshort{ilt}), thrombus for short, and wall \textit{calcification}.
The former is the stratification of several blood particles in the inner layers, the latter is the accumulation of calcium in the aortic wall responsible of its stiffening.
Both negatively affect the aorta changing its physiological behavior by modifying the \textit{lumen}'s structure, the region where the blood flows, with bumps and irregularities.

Since vessel's morphology plays a major role in local haemodynamics which, in turn, determines the development of the \acrshort{aaa} \cite{1_dinamica_01},  \cite{1_intro_07}, in this study we focus on the aortic lumen. Hence we are only interested in the information directly associated to the shape of the lumen. For this reason, calcium formations which could occur on the external part of a thrombus will be disregarded.

\section{Data}\label{sec:data}

In this section we outline our complex pipeline for data gathering and pre-processing.

\subsection{CTA scans}

The most used and effective method for anatomy visualization is \acrlong{cta}.
\acrshort{cta} is used to get a visualization of the human body and is realized with emission of X-rays through a tomograph.
A \acrshort{cta} scan consists of a 3-dimensional image of a region of interest in the form of a volume composed by voxels with different shades of grey. Each shade is associated to a different tissue, making it possible to distinguish not only between fat, bones and muscle but also artery, \acrlong{ilt} (\acrshort{ilt}) and calcification - see also \Cref{fig:CTA}.

The data considered in this work consist of 48 \acrshort{cta} scans of the abdominal region, 24 of which picture a pathological aorta. To secure data consistency, the following criteria have been applied:
\begin{itemize}
    \item \acrshort{aaa} greater than $3.5 cm$ in diameter;
    \item infrarenal \acrshort{aaa}, not extended above the renal arteries, and presence of the aortic neck;
    \item \acrshort{cta} with high resolution and contrast, to assure high quality data;
    \item non-minor and non-pregnant patients.
\end{itemize}

All the \acrshort{cta} scans are provided by Ospedale Maggiore Policlinico. Patients with healthy aortas have been collected from the Policlinico's archive while the patients affected by \acrshort{aaa} come from the Vascular Surgery Unit. For privacy reasons, all data are anonymized.

\subsection{Segmentation Pipeline}

\begin{figure}
    \centering
    \includegraphics[width=0.9\textwidth]{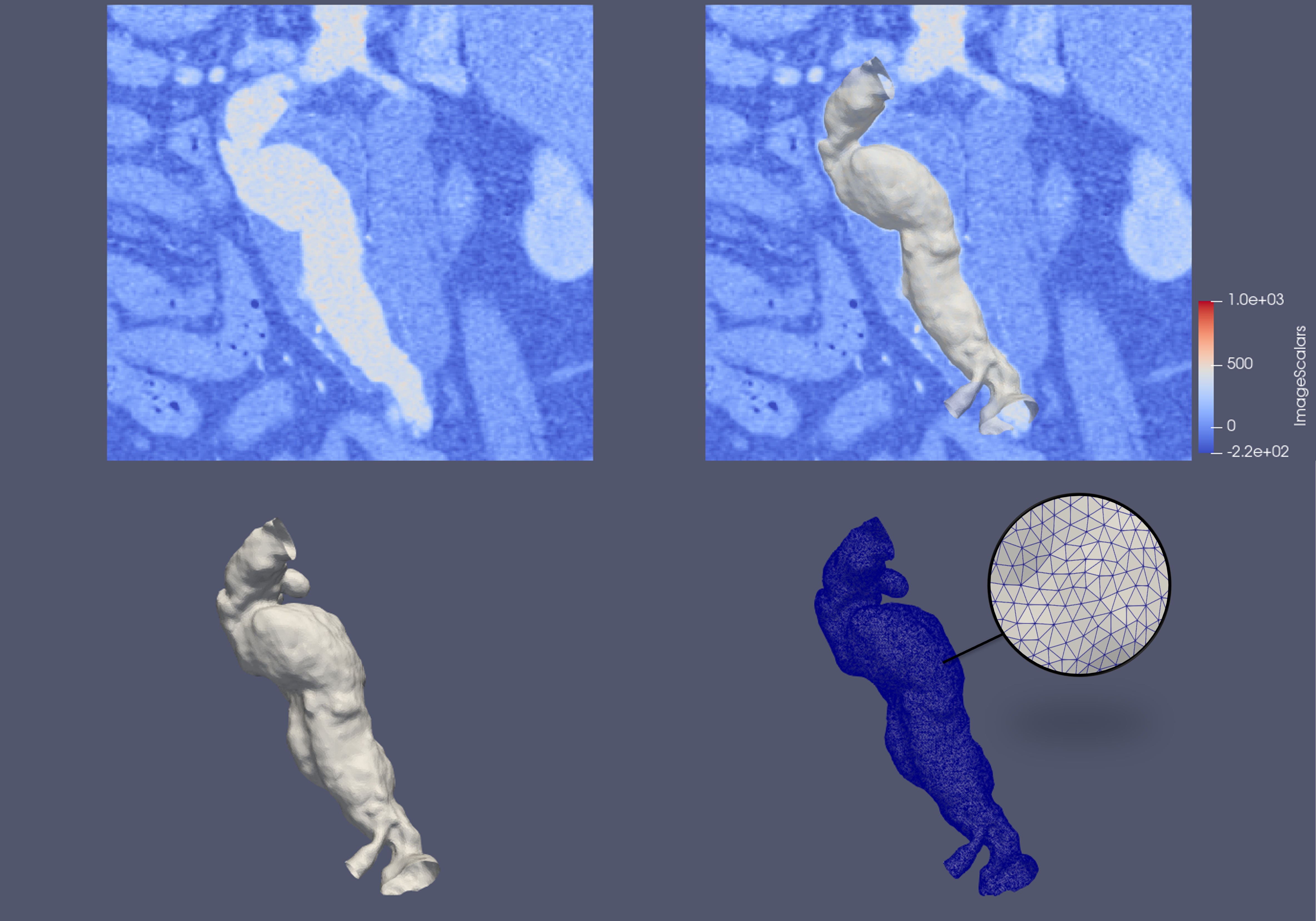}
    \caption{A schematic view of the segmentation process, starting from the CTA scan and obtaining the final mesh representing the portion of interest of the Aorta and the iliacs.}
    \label{fig:segmentation}
\end{figure}

The primary function of a \acrshort{cta} scan in this study is to enable the segmentation of the aortic lumen, rather than for visualization purposes.
Aortic segmentation refers to the process of obtaining a three-dimensional reconstruction of the lumen via a finite number of points, as visually described in \Cref{fig:segmentation}. The product of the segmentation is a mesh, consisting of a set of vertices, edges and cells that form a net of triangles. 
Segmentation is made possible thanks to a pipeline provided by \textit{Moxoff}\footnote{https://www.moxoff.com/en/}.

The tract of the vessel considered in this study ranges between the distal renal artery and the common iliac arteries, including the area eventually affected by the \acrshort{aaa} and iliac aneurysms, using manually selected points - see also \Cref{fig:CTA}. 
The pipeline used to perform the segmentation relies mainly on an edge-based technique with the support of thresholding and region-based methods to improve performances and results. Following \cite{3_ranges}, thresholding makes an initial guess of the points belonging to the lumen. Then region-based and edge-based techniques \cite{3_region} help with recognition of the rounded shape of the vessel, when projected on the coronal plane.
In this way, a label is applied to each of the voxels of the \acrshort{cta} scan creating a volume whose surface is extracted and triangulated.

The resolution of the segmentation (i.e. the average length of the edges) is $1.0mm$, empirically chosen to balance the trade off between accuracy and computational costs.

A by-product of the segmentation is the \emph{centreline}. Roughly speaking, the centerline is a continuous curve in space that represents the center of the vessel lumen and makes for a powerful descriptor of its shape. Among the several methods proposed in the literature for its identification, we use the algorithm provided by Antiga et al. (2008) within VMTK, the Vascular Modeling Toolkit\footnote{http://www.vmtk.org}, which generates a mathematical object stable to perturbations of the vessel's lumen surface. Identification of the centerline is of greatest importance for singling out points on the lumen surface considered as peripheral.


\section{Object Representation through Persistent Homology}
\label{sec:PH}

We now describe the mathematical objects used to represent the atoms of our statistical analysis, that is to represent the meshes obtained from segmentation of the CTAs of the patients.
The key element that drives the upcoming construction is a function defined on a mesh, called \emph{filtering} function, with the property that the morphological features we are interested in 
appear as local minima and local maxima of this function. In particular, the filtering function should capture inward and outward bumps of a blood vessel, so it must be based on some sort of radial distance.

\subsection{Filtrations and Homology}\label{sec:filtration}

Consider a triangular mesh $K\subset \R^3$ as the union of the set of its triangles, $K^2$, the set of its edges, $K^1$, and the set of its vertices $K^0$. This data amounts to a \emph{simplicial complex}
\cite{munkres2018elements} which is the combinatorial starting point of persistent homology.
Every element in $K$ is called a simplex: \emph{0-simplex} if it is a vertex, \emph{1-simplex} if it is an edge and \emph{2-simplex}, if it is a triangle.

To define the \emph{radial filtering function} $f$, we start with a function $f:K^0\rightarrow \R$ defined on the vertices of the mesh, and we suitably extend it to the whole $K$.
Let $\gamma\subset \R^3$ be the centerline of the blood vessel. For every vertex $v\in K^0$ define:
\[
f(v)=\min_{p\in \gamma} \parallel v-p\parallel,
\]
that is, the distance of the vertex \(v\) from the centreline. 
For every $\sigma\in K^1 \bigcup K^2$, being it an edge or a triangle, we extend $f$ as follows:
 \[
 \textstyle f(\sigma)= \max_{v\in \sigma \cap K^0} f(v). 
 \]
 This extension is motivated by the following facts: it is easy to handle computationally and, by standard topology results \cite{edelsbrunner2022computational}, it is equivalent to the piecewise linear extension on the mesh of the $f$ defined on its vertices.

We now use $f:K\rightarrow \R$ to order the vertices, edges and triangles of $K$ into a filtration of simplicial complexes. For \(t \in \R,\) consider the simplicial complex $K_{t}$ defined by the sublevel set 
\[
K_{t} = f^{-1}((-\infty,t]).
\]
Since \(f\) can assume only a finite number of distinct values, let them be \(t_0<t_1<...<t_m\) to obtain the filtration \(\{K_{t_i}\}\) of simplicial complexes, such that
\[
K_{t_0} \subset K_{t_1} \subset \ldots \subset K_{t_m} = K.
\]

\begin{figure}
  \begin{subfigure}{0.45\textwidth}
    \centering
    \includegraphics[width=0.95\textwidth]{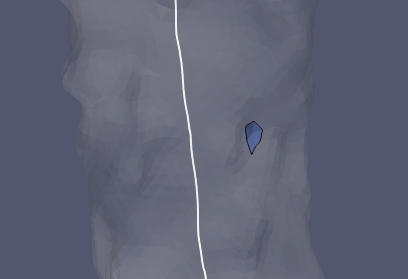}
    \caption{A path connected component created by an inward bump, in the sublevel set filtration defined in \Cref{sec:filtration}. Path connected components created by other inward bumps merge with each other at saddle points or local maxima.}
    \label{fig:min}
  \end{subfigure}
  \begin{subfigure}{0.45\textwidth}
    \centering
    \includegraphics[width=0.95\textwidth]{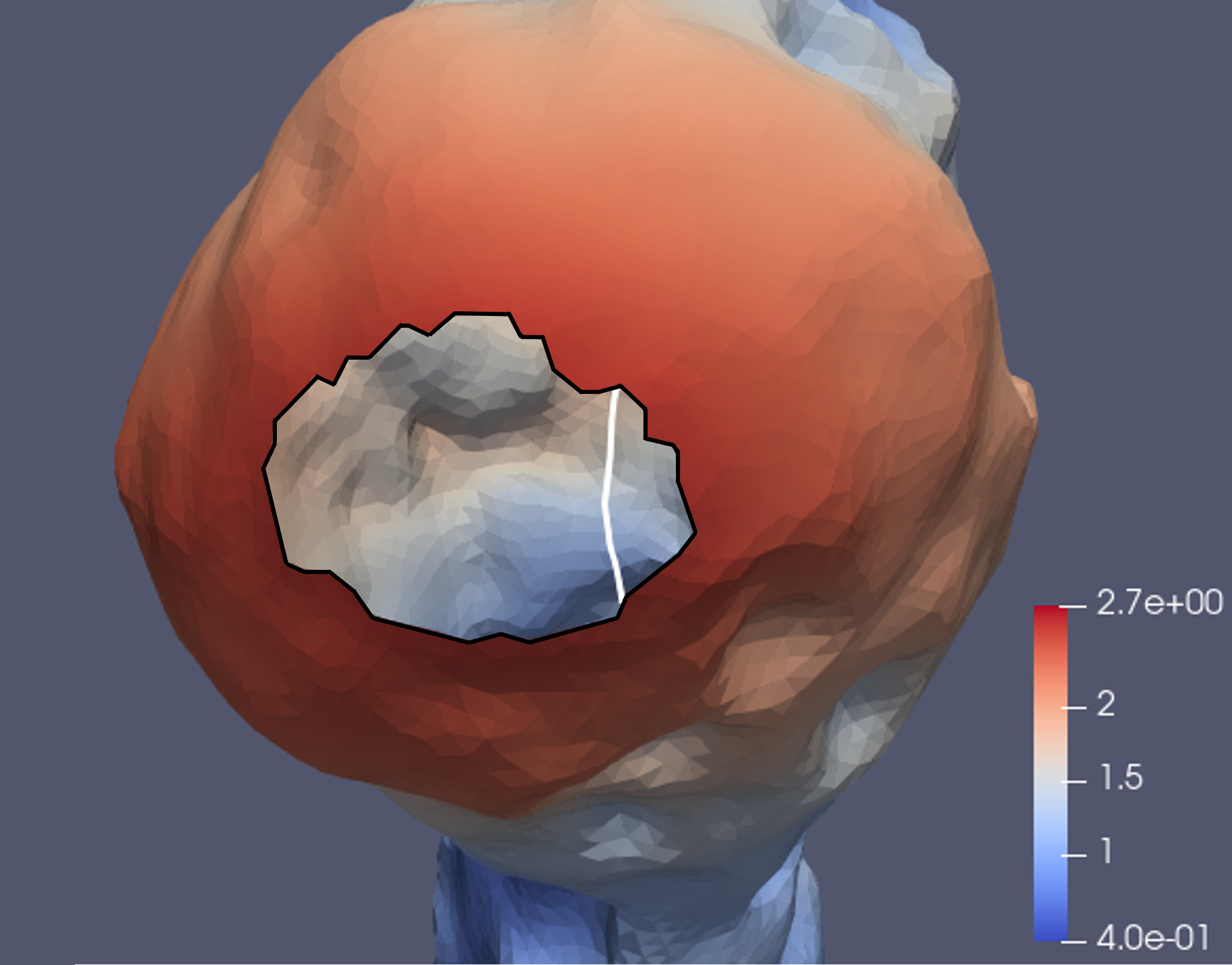}
    \caption{A loop created by an outward bump, in the sublevel set filtration defined in \Cref{sec:filtration}. When the whole bulge is contained in the filtration, the loop disappears.}
    \label{fig:max}
  \end{subfigure}

    \centering
    \begin{subfigure}{0.2\textwidth}
    \centering
    \includegraphics[width=\textwidth]{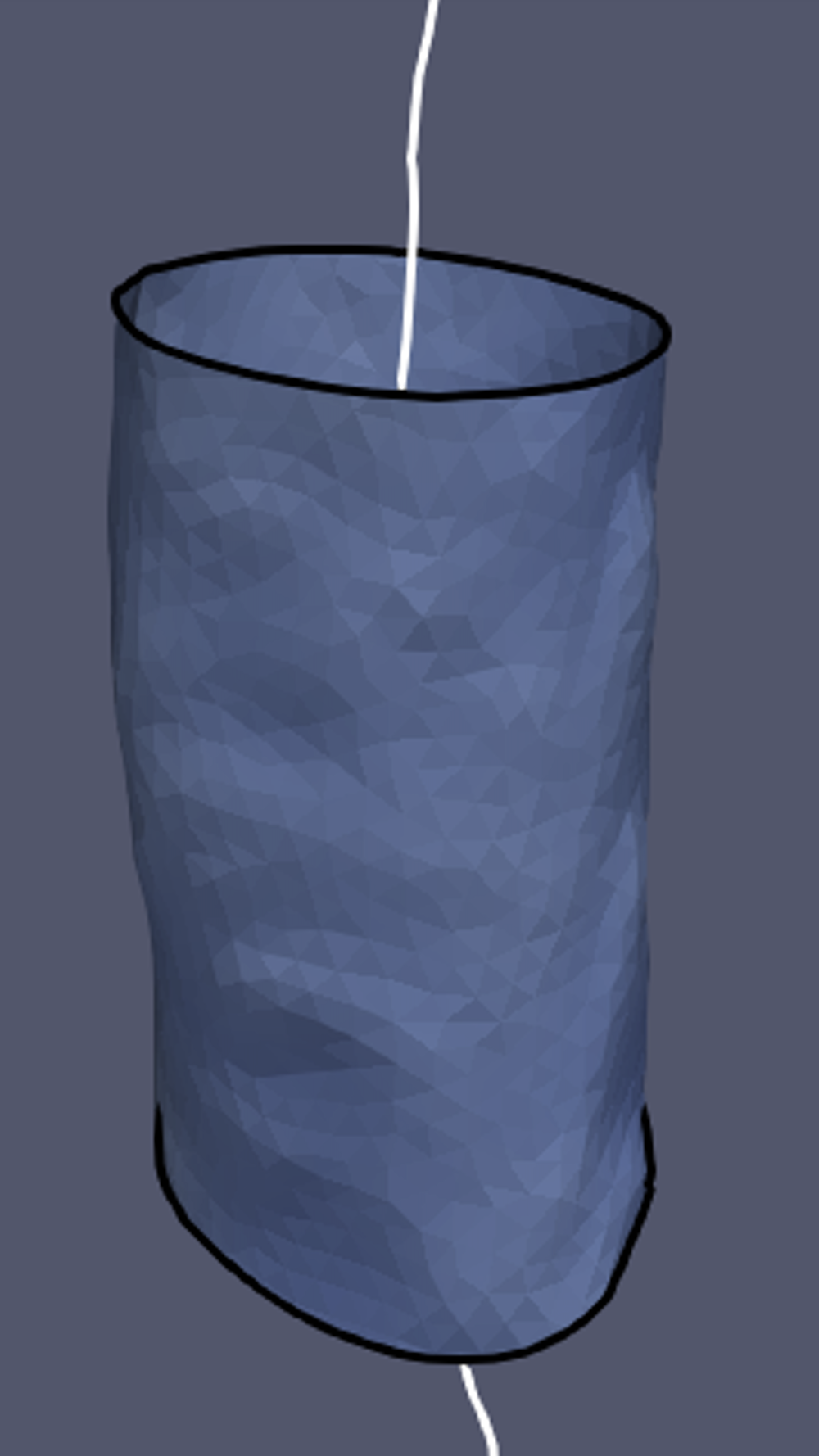}
    \caption{Loops generated by the tubular structure of the mesh.}
    \label{fig:typeloops}
    \end{subfigure}
    \begin{subfigure}{0.75\textwidth}
    \centering
    \includegraphics[width=1\textwidth]{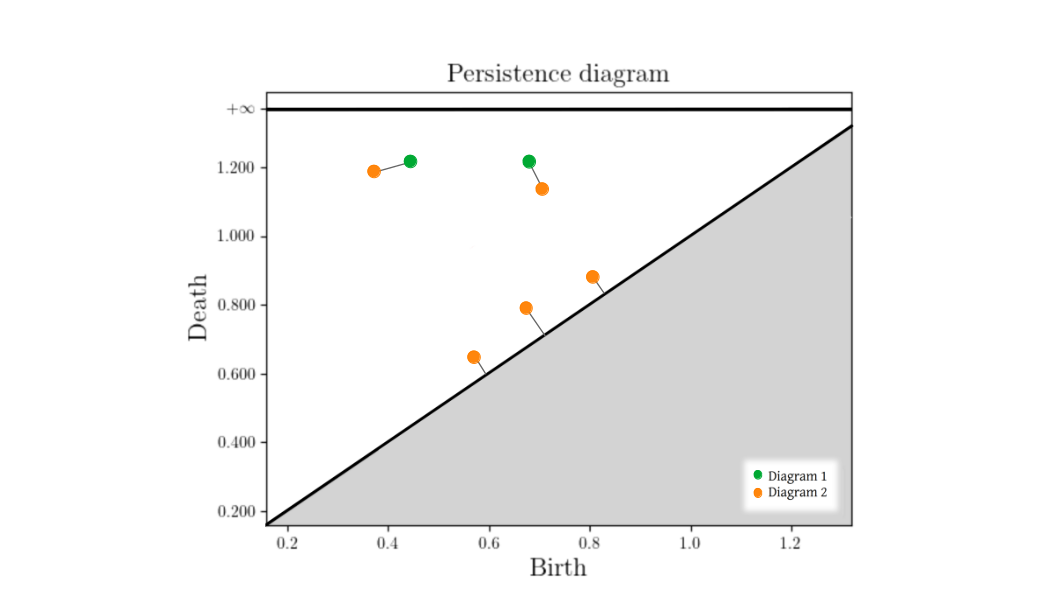}
    \caption{Two persistence diagrams and the optimal matching between them giving the bottleneck distance.}
    \label{fig:metric}
    \end{subfigure}
    
    \caption{Path connected components, loops and persistence diagrams.}
  \label{fig:min_max}

\end{figure}

Some pivotal observations are in order; they can be visualized by looking at \Cref{fig:min_max}.

First we define the path connected components of \(f\).
An inward bump of the vessel corresponds to a local minimum of $f;$ when it appears in the filtration 
\(\{K_{t_i}\}\), a new path connected component is \virgolette{born} in the sublevel sets of \(f;\) see \Cref{fig:min}.  
That path connected component persists until it is merged with other path connected components, born at lower local minima. Visually, for this to happen, one needs that the whole inward bump generated by the local minimum is added to the filtration, so that different path connected components can meet at local maxima or saddle points. An outwards bump, instead, produces a loop which surrounds its boundary  - see \Cref{fig:max}. As the filtration value increases and the simplicial complexes grow, this loop, which encloses the local maximum of the outward bump, becomes smaller and smaller, eventually superimposing to the local maximum itself, and \virgolette{dying}  - see \Cref{fig:max}. There is also another way in which loops are created, as showcased by \Cref{fig:typeloops}: the tubular structure of the blood vessels allows also loops that go around the centerline. Two such loops are equivalent if they can \virgolette{slide} on the mesh, superimposing one with the other - like the top and the bottom black loops in \Cref{fig:typeloops}.


Thus, the mesh is transformed through \(f\) into the filtration of simplicial complexes $\{K_{t_i}\};$ identifying along the filtration \virgolette{births} and \virgolette{deaths} of path connected components and loops, characterizes the blood vessel in terms of the irregularities of its lumen. 

Topologically speaking path connected components and loops are captured, respectively, by homology in dimension 0 and 1, which we now formally and briefly define. For more details see \cite{edelsbrunner2022computational}.

For notational coherence with the literature, we refer to (oriented) triangles and edges in a simplicial complex via the set of their vertices enclosed by squared brackets:  e.g. $[x_0,x_1,x_2]$ for a triangle, i.e. a 2-simplex, and $[x_0,x_1]$ for an edge, i.e. 1-simplex. Given a simplicial complex $K_{t_i}$, for $n = 0,1,2$ we generate the vector space $\ch_n(K_{t_i})$ over the field $\mathbb{Z}_2=\{0,1\},$
by considering the set of all finite formal sums of the \(n\)-simplices belonging to $K_{t_i}$.
For $n =1,2,$ the boundary operators $\partial_n:\ch_n(K_{t_i})\rightarrow \ch_{n-1}(K_{t_i})$ are then defined by setting
\[
\partial_n(\sigma)=\sum_{i=0}^n \sigma_{-i},
\]
when \(\sigma\) is an \(n\)-simplex, and by extending linearly to the whole vector space $\ch_n(K_{t_i});$ here \(\sigma_{-i}\) is the (oriented) \((n-1)\)-simplex obtained from \(\sigma\) by deleting its i-th vertex -- e.g. \([x_0,x_1,x_2]_{-1}=[x_0,x_2].\)  The boundary operator \(\partial_0\) maps \(\ch_1(K_{t_i})\) in the trivial vector space whose only element is 0.
We can now introduce the spaces of $n$-boundaries and $n$-cycles of $K_{t_i}$: $\Z_n(K_{t_i}) = \ker(\partial_n)$ are the $n$-cycles and $\B_n(K_{t_i})=\im(\partial_{n+1})$ are the $n$-boundaries of $K_{t_i}$. Finally, we define the $n$-dimensional simplicial homology groups $\h_n(K_{t_i})= \Z_n(K_{t_i})/\B_n(K_{t_i})$. 
Note that the quotient $\Z_n(K_{t_i})/\B_n(K_{t_i})$ is well defined since $\partial_n \circ \partial_{n+1} =0$.
In particular, linearly independent vectors in $\h_1(K_{t_i})$ are independent loops which are not filled by triangles (see \Cref{fig:min_max}, right), and
linearly independent vectors in $\h_0(K_{t_i})$ are points which are on different path connected components (see \Cref{fig:min_max}, left). Elements in $\h_n(K_{t_i})$
are referred to as homology classes or (equivalence) classes of $n$-cycles. Tracking down the evolution of $\h_1(K_{t_i})$
and $\h_0(K_{t_i})$ along the filtration \(\{K_{t_i}\}\) returns the information on the vessel shape we are going to explore in this paper.

\subsection{Normalization}\label{sec:normalization}

One further step is needed to ensure a fair comparison between filtrations obtained from vessels of different patients. 

As it is often the case with shape analysis, the magnitude or the size difference between patients can act as a confounding factor to the point of overshadowing the shape variability, which is the one we are interested in. For instance, different patients can have different healthy dimensions for the aorta and for the iliacs, making it harder to compare them by contrasting the filtrations and the homology groups generated by the their respective filtering functions.
To avoid this potential bias in the analysis, we resort to a normalization process, expressing all distance values relative to the healthy size of the patient's aorta, i.e. the average neck radius. We recall that the aortic neck is in fact the first healthy tract of the aorta, right below the distal renal artery. Its diameter is equal to an healthy aortic diameter, it does not have high variability and it summarizes the size factor of the aorta, in accordance with the scientific literature \cite{5_AMQGC}.
So, for each patient, the filtering function values have been divided by its average neck radius, obtained as reported in \Cref{app:neck_radius}. 

On top of that, since different portions of the blood vessel are characterized by a different relative radius, with respect to the neck's radius, we can in fact relate the normalized filtering function values to morphological features of the aorta, see \Cref{sec:read_PDs}.

\subsection{Persistence Diagrams}

Looking at the homology groups obtained from a filtration is not practical for applications, but there is a very interpretable and concise summary of the information contained in the sequence of $\h_n(K_{t_i})$, called the \emph{persistence diagram} in dimension $n$. These are the objects that we are going to use to represent data.

A persistence diagram is a finite collection of points in $\R^2$ - also called \emph{persistence pairs} -, with every point $p=(b,d)$ representing the birth and the death of a persistent homology class. In particular, the birth of an homology class is the first \virgolette{time \(t\)}  it appears along the filtration, while the death time happens when the class merges with another class born earlier. For example: a local minimum of the filtering function, indicative of a bump in the vessel, induces a path connected component which is born at the value of the minimum, \Cref{fig:min}; such path connected component carries on, or \emph{persists}, along the filtration, and dies when it merges with a path connected component born earlier, therefore associated to a lower local minimum. The absolute difference between the times of birth and death is clearly related to the prominence of the bump, and is called \emph{persistence} of the homological feature.
Note that points can appear multiple times in the same persistence diagram. 

\subsection{Populations of Persistence Diagrams}

Lastly, it is possible to
compare and analyize the topological information carried by different persistent diagrams by defining suitable metrics, satisfying formal stability results; for an extensive review see, for instance, \cite{edelsbrunner2022computational}. 
It is thus possible to quantify the dissimilarity of two different diagrams and perform classification, clustering and dimensionality reduction on populations of diagrams.

To simplify the upcoming formulas, let 
\[
D = \{(b_1,d_1),\ldots,(b_n,d_n)\mid n\in \mathbb{N}, b_i,d_i \in \R \cup\{\infty\}, b_i < d_i\}\cup\{(b,d)\mid b,d \in \R, b=d\}. 
\]
represent a persistence diagram.
Given two diagrams $D_1$ and $D_2$ and $p\geq 1$, the $p$-Wasserstein distance between them is:

\begin{equation}
\label{p_Wasserstein_dist}
W_p(D_1,D_2) = \left( \inf_\gamma  \sum_{x\in D_1} \parallel x-\gamma(x)\parallel^p_\infty \right)^{1/p}
\end{equation}
where the \(\inf\) is taken over all bijections  $\gamma$ between diagrams $D_1$ and $D_2$. In other words we measure the distances between the points of the two diagrams, pairing each point of a diagram either with a point on the other diagram, or with a point on $y=x$ (see \Cref{fig:metric}). Each point can be matched once and only once. The minimal cost of such matching provides the distance. The case $p=\infty$ is usually referred to as the bottleneck distance and has the following form:
\[      
W_\infty(D_1,D_2) = \inf_{\gamma} \sup_{x \in D_1} \parallel x-\gamma(x)\parallel_\infty. 
\]
Note that, to compare two persistence diagrams, we do not need to establish a relationship between the generating meshes. In particular this means that no alignment or registration is needed to proceed with the analysis; this is the ancillary source of variability between patients we are not interested in.

Finally, let us mention an important stability result \cite{cohen2005stability}. Given two filtering functions $f,g:K\rightarrow \R$ and their respective persistence diagrams $D(f),D(g)$: 
\[
W_\infty(D(f),D(g)) \leq \parallel f-g\parallel_\infty,
\]
meaning that persistence diagrams are a faithful representation of the functions, in terms of the sup-norm. 


Closing this section on persistent homology, we point out the existence of representations alternative to persistence diagrams, for instance persistence landscapes \cite{landscapes}, persistence images \cite{pers_img}, persistence silhouettes \cite{silhouettes}, accumulated persistence functions \cite{biscio2019accumulated}. All of these come with their own properties and stability results and can be used for analyses which are possibly more refined or more tailored to the application of interest.

\section{Reading a persistence diagram}\label{sec:read_PDs}
To understand the descriptive power of the persistence diagrams obtained with the pipeline illustated in the previous section, a parallel reading of the original mesh and the associated diagrams has been made for every patient in the study, connecting the most important aortic wall features - such as AAA, calcifications and thrombus - to the diagram points. 

All the results contained in the upcoming subsections have been manually verified by simultaneously looking at the persistence diagram, the simplicial complexes and the \acrshort{cta} scans, with the support of the collaborating clinicians.

\subsection{The Effect of Normalization}
\label{sec:normalization_effect}
We have anticipated in \Cref{sec:normalization} that normalization is instrumental for comparing different patients since it allows for the appraisal of the distance from the centerline when a persistence pair is created or ceases to exist, relatively to the size of the mean aortic neck radius.
Normalization is, in fact, also significant for the identification of the portion of the aorta where the change in homology occurs, locating it on a specific section of the vessel. 
Note that the diameter of the vessel differs along the aorta, especially after the aortic bifurcation.
In fact, the diameter of a healthy abdominal aorta is tipically  $17.5 \pm 2.1 mm$, larger than that of an healthy iliac artery, $10.85 \pm 1.69 mm$; see \cite{7_dimensions_1, 7_dimensions_2}. 
Although significant differences in the dimension of the aorta for males and females occur, the proportion between the diameter of the main aorta and that of the iliac branches are similar.
Thus:

\begin{itemize}
    \item when an homology class is born (or dies) at a value close to the mean neck's radius, i.e. a value close to 1, the change must be located at the same distance from the centerline as the aortic mean neck's radius, hence on the healthy portion of the aorta. The only exception being an iliac aneurysm, whose presence will be discussed later in \Cref{sec:natablepairs};
    \item in the cohort of the patients considered in this study, the diameter of healthy iliacs is around 40-70 \% of the respective aortic mean neck's radius.
Thus, changes in homology located on the iliac arteries occur when the normalized distance assumes values in the range of $[0.4-0.7]$;
\item lastly, patients affected by \acrshort{aaa} can instead have a more variable range, up to 7 and more in the most concerning cases. 
\end{itemize}

\begin{figure} 
\centering
\includegraphics[width=0.8\textwidth]{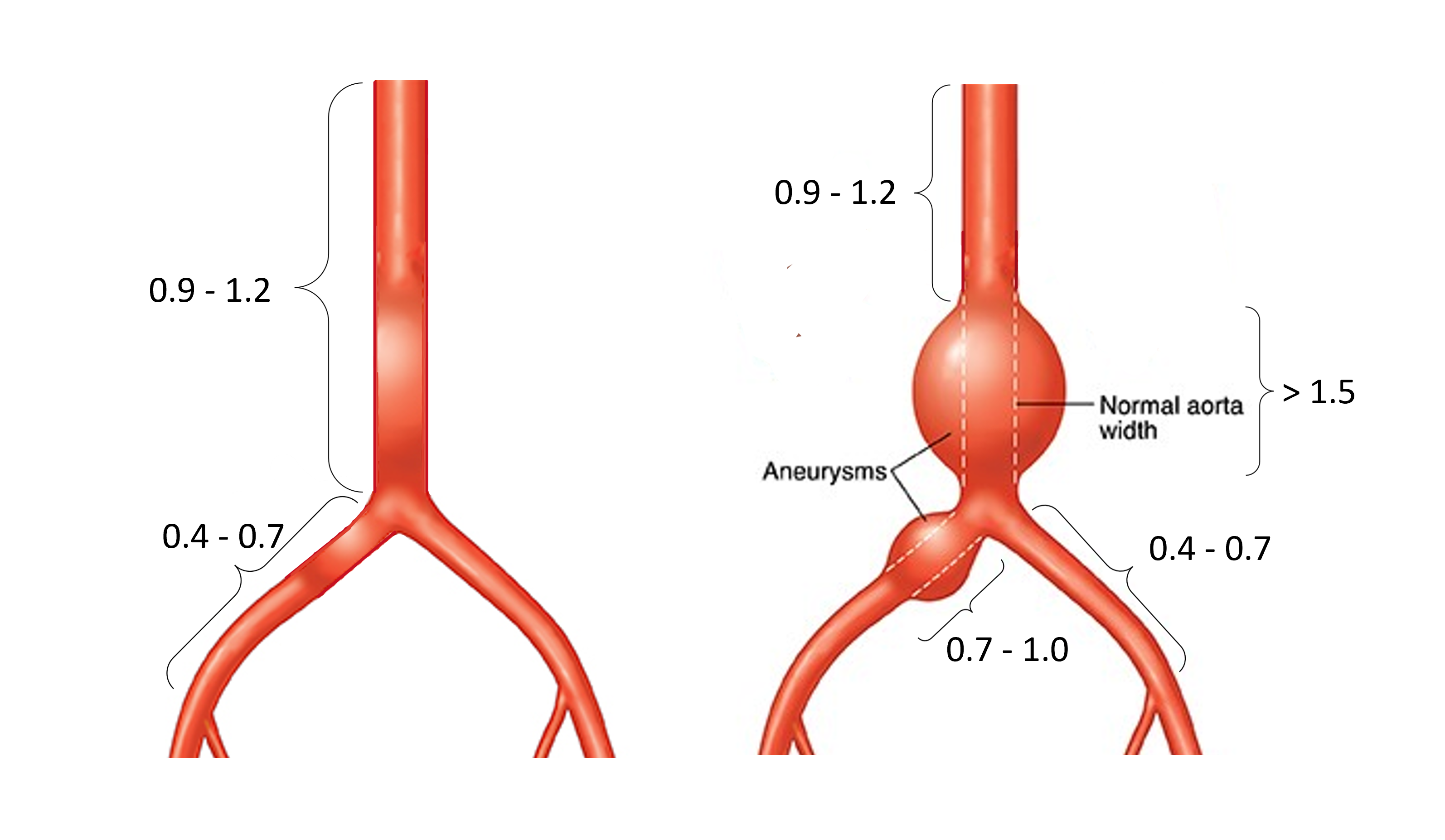}
\caption{A schematic view of the aorta, with the normalized distances from the centerline made explicit. Image modified from 
\protect\url{https://prescrivere.blogspot.com/}.}
\label{fig:rayscheme}
\end{figure}

\Cref{fig:rayscheme} depicts a schematic visualization of these facts.

\subsection{Notable Pairs}\label{sec:natablepairs}

By combining the information given by birth, death, persistence and type of pairs of points in the persistence diagram, it is possible to identify \emph{notable pairs}, that is persistence pairs shared among patients with similar characteristics associated with a specific aortic region.

\subsubsection{Structural Pairs}

The first class of points we consider are those associated to an infinite persistence, i.e. such that $d=\infty$. These points correspond to features which are born at a certain filtration value, but which never die, and so persist up to $\infty$.
They indicate structural features shared by all aortas. In particular, each filtration coming from a correctly segmented aorta must have exactly three points with infinite persistence: one associated to 0-cycles and two associated two 1-cycles.

\begin{figure} 
    \centering
    \includegraphics[width=\textwidth]{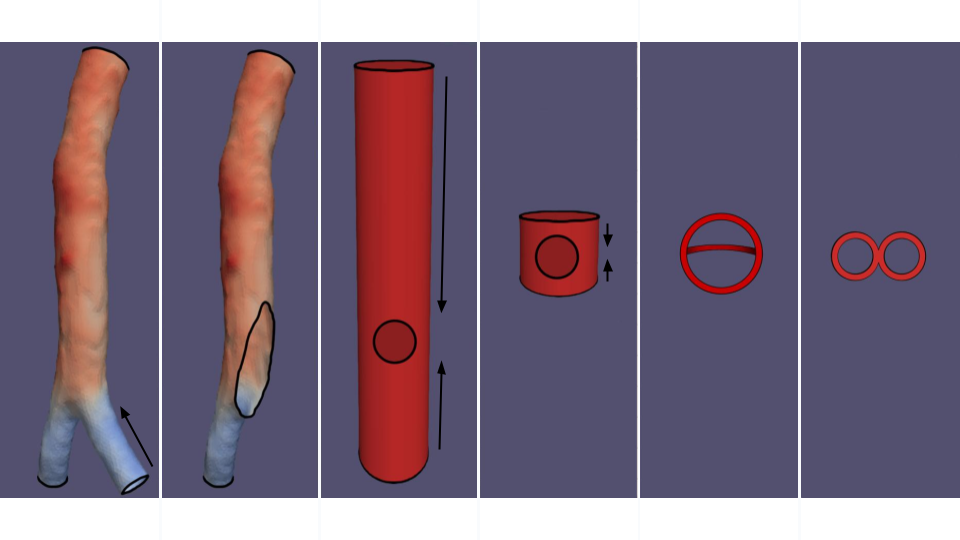}
    \caption{All the objects in this figure are homotopy equivalent.}
    \label{fig:quadtree2}
\end{figure}

Indeed, the mesh representing the aorta should be \emph{homotopy equivalent} 
\cite{edelsbrunner2022computational}  to a finite cylinder with a hole, which, in turn, is equivalent to an eight figure: two circles glued in one point, see \Cref{fig:quadtree2}.
Roughly speaking, two objects are homotopy equivalent if, starting from one, it is possible to obtain the other using stretching and bending but not tearing. Homotopy equivalence preserves homology and so each mesh should feature
 a connected component with two independent 1 dimensional loops as can be seen in \Cref{fig:quadtree2}.

These three points, representing the most persistent homological features, can also be used as indication of the correct representation provided by the mesh itself, proving the deep connection between the persistence diagram and the aorta. In fact, if a persistence diagram shows additional points with infinite persistence, this signals an error during the segmentation phase, such as the creation of non-connected points or the presence of artificial holes in the mesh.

\subsubsection{Iliac aneurysm}\label{sec:iliacs}

\begin{figure} 
  \centering
  \begin{subfigure}{\textwidth}
    \centering
    \includegraphics[width=0.8\linewidth]{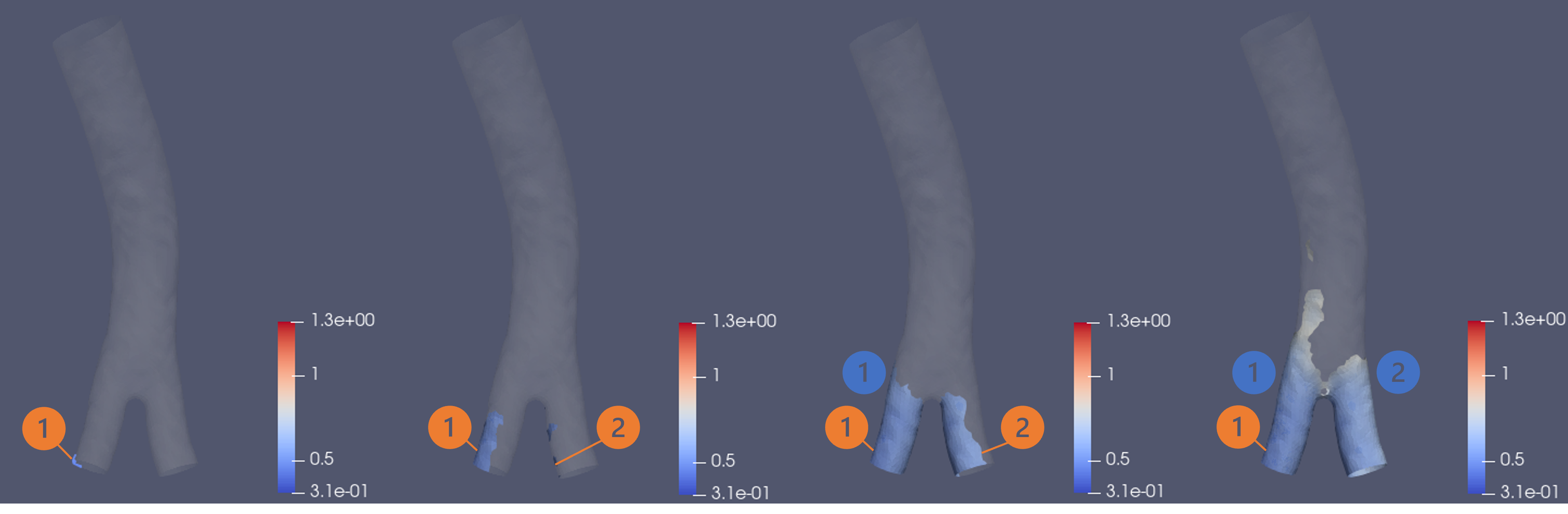}
    \caption{Initial steps of the sublevel set filtration of a patient with healthy iliacs, showing the births of the iliacs-related persistence pairs: first appear the two local minima of the radial distance located on the two iliacs - orange (1) and (2) - and then the two loops going around the tubular structure of the iliacs - blue (1) and (2).}
    \label{fig:subfig1}
  \end{subfigure}
  \begin{subfigure}{\textwidth}
    \centering
    \includegraphics[width=0.8\linewidth]{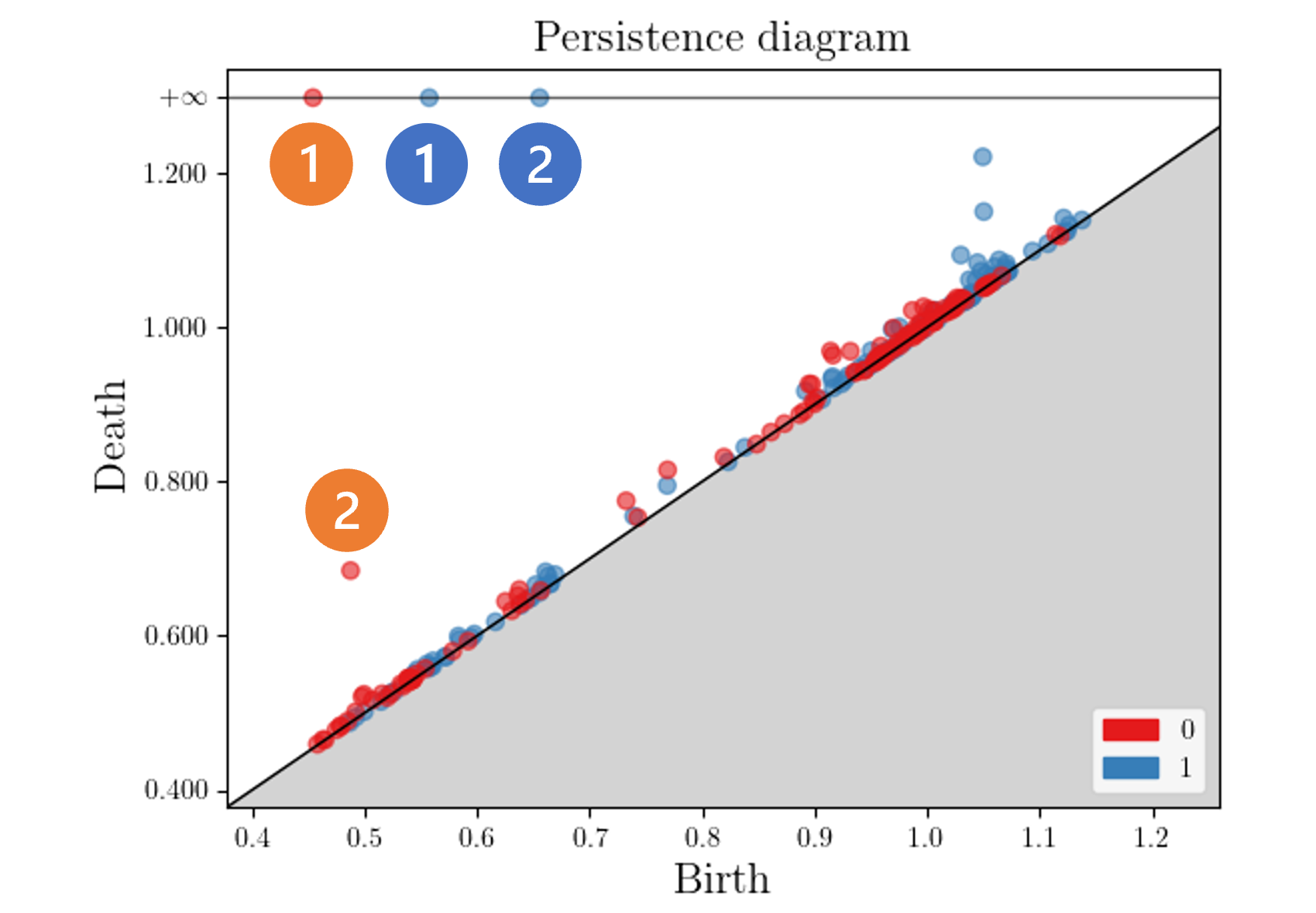}
    \caption{The persistence diagram, in dimension 0 and 1, of an healthy patient, with highlighted the persistence pairs related to the iliacs, whose labels are coherent with the ones in \Cref{fig:subfig1}.}
    \label{fig:subfig2}
  \end{subfigure}

  \caption{A visualization of the mesh of a healthy aorta with the birth and death of cycles related to the iliacs highlighted in the associated diagram.}
  \label{fig:healthyexample}
\end{figure}

\begin{figure}
  \begin{subfigure}{0.6\textwidth}
    \centering
    \begin{subfigure}{\textwidth}
      \centering
      \includegraphics[width=0.75\textwidth]{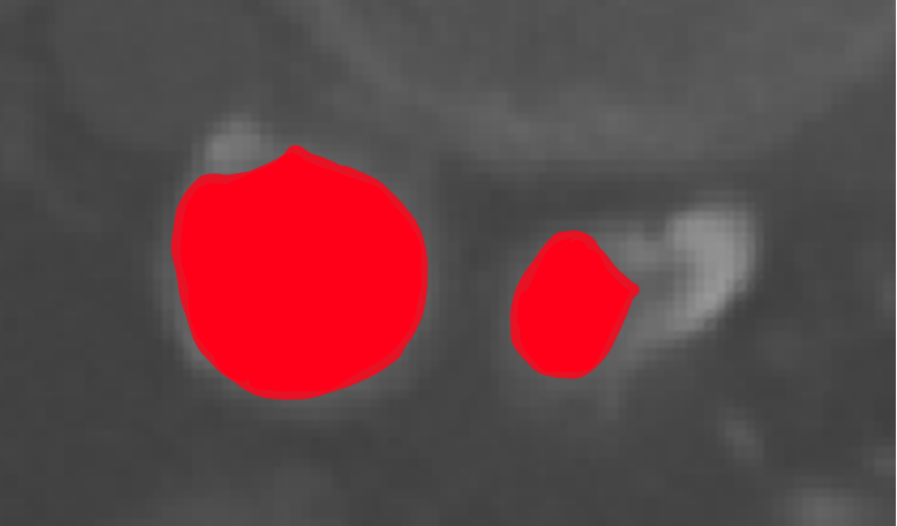}
      \caption{CTA scan slice of iliacs presenting calcification in the right iliac and aneurysm in the left. The red portion shows the lumen restriction created by calcifications and the bulge created by the aneurysm.}
      \label{fig:immagine1}
    \end{subfigure}
    \vspace{10pt} 
    \begin{subfigure}{\textwidth}
      \centering
      \includegraphics[width=\textwidth]{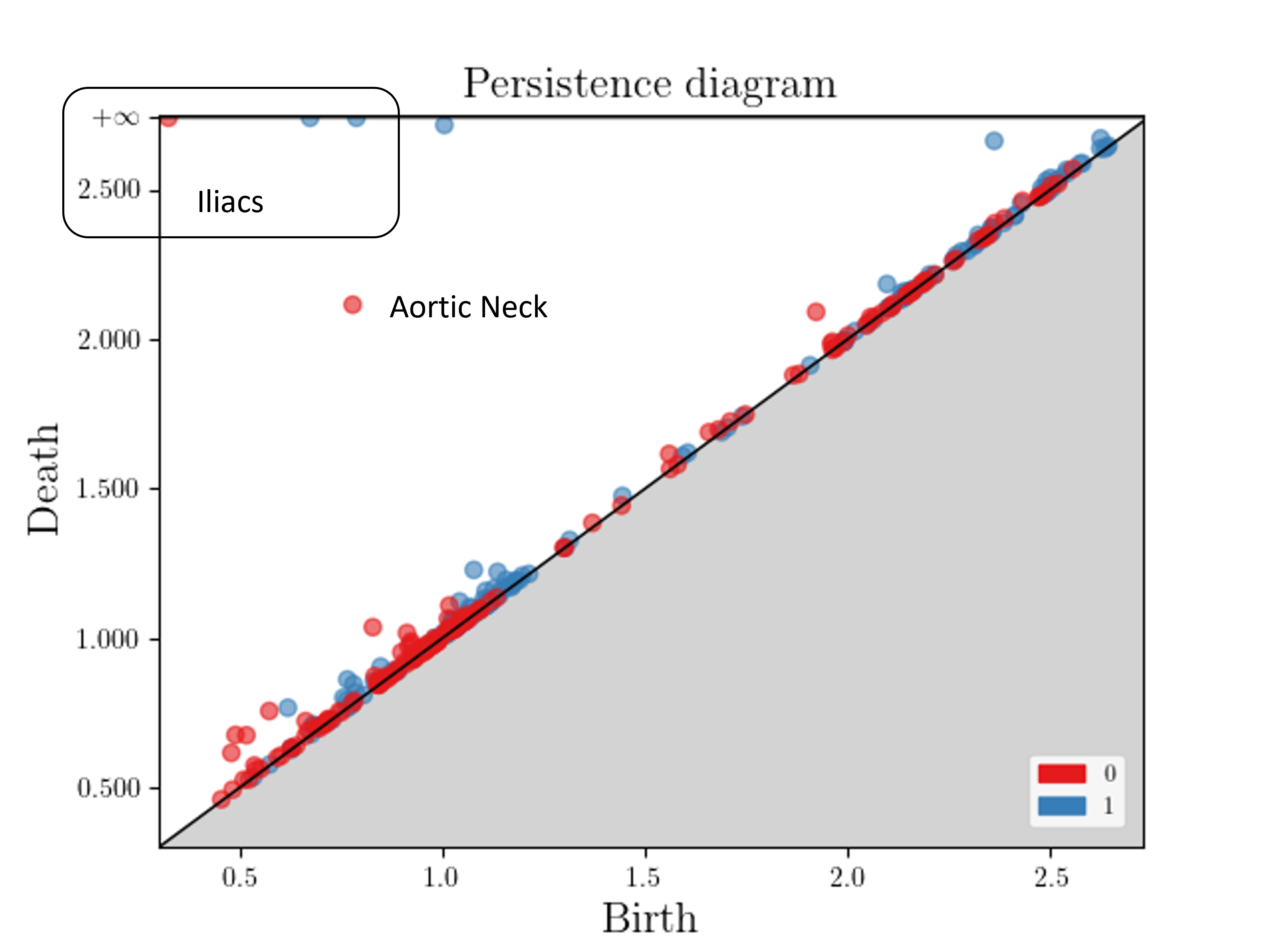}
      \caption{The persistence diagram of a patient with an aortic aneurysm and an iliac aneurysm. The lack of an early $0$-cycle with high persistence is visible. As is the presence of one class of high persistence 1-cycles created by the AAA.}
      \label{fig:iliacaneurysmdiagram}
    \end{subfigure}
  \end{subfigure}
  \begin{subfigure}{0.35\textwidth}
    \centering
    \includegraphics[width=\textwidth]{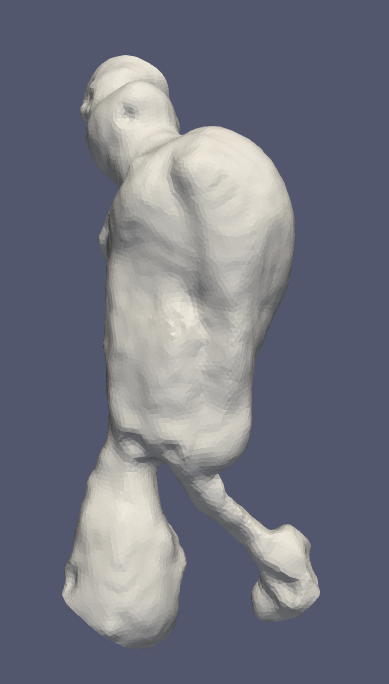}
    \caption{Mesh of the lumen of a patient being affected by calcifications and aneurysm in the iliacs: the left one presents a bulge created by the aneurysm, while the right one is severely occluded by calcifications.}
    \label{fig:immagine3}
  \end{subfigure}

  \caption{\acrshort{cta} scan slice, mesh and persistence diagram of the same patient with a calcified (left) and aneurysmatic (right) iliac.}
  \label{fig:iliacs}  
\end{figure}

Now we tackle the problem of identifying iliac aneurysms; the reader should refer to \Cref{fig:healthyexample} and \Cref{fig:iliacs} to get a visual description of the content of this section. 

Aneurysms can affect one or both the iliac arteries, in the same way as they affect the abdominal aorta, resulting in a stretch of the wall - see \Cref{fig:immagine3}. An iliac aneurysm can also be associated with calcification (as in \Cref{fig:iliacs}) and thrombus, just as the \acrshort{aaa}.

When the iliacs have comparable diameters there are two connected components born at similar values, resulting in two classes of 0-cycles that later merge together - as in \Cref{fig:healthyexample}. 
The single class of 0-cycles with infinite persistence represents the closest point of the mesh to the centerline, the very first point appearing in the filtration. This class is always born at the iliacs, given their smaller radiuses. Moreover, the other iliac's first point induces the birth of a secondary early class of 0-cycles with high persistence. These components will be merged together as the radius value increases (see point \virgolette{(1)} and \virgolette{(2)} in \Cref{fig:healthyexample}). For these reasons, each persistence diagram, except the ones with iliac aneurysm, have a high persistence point with birth coordinate given by the radius at which the second component is born and death coordinate given by the merging radius. 

When an iliac aneurysm is present, it significantly deforms the lumen and so the minimal radius of the aneurysmatic iliac is greatly increased - see \Cref{fig:iliacs}. As a consequence the first point of the second iliac appears much later in the filtration, often after other points at the bifurcation region. Thus, instead of having two connected components born at similar times, we have a component that is expanding until it covers both iliacs. As a consequence, the absence of this second high persistence point in dimension $0$ is a primary indicator for an iliac aneurysm. 

Similarly, any irregularity in the iliacs, inflating their lumen, may have the effect of delaying the birth of $1$-cycles with infinite persistence. 
For instance, not having two classes of 1-cycles with infinite persistence born at roughly the same times (i.e. having roughly the same diameters) implies that one iliac is inflated. The only reasonable cause for that inflation is the presence of an aneurysm. Thus, if the two classes of $1$-cycles with infinite persistence have a pronounced difference between their birth coordinates or have birth coordinates which are high (note that a $1$-cycle is always going to appear at radius $1$, around the neck), this indicates the presence of one or more aneurysms in that region of the blood vessels.

\subsubsection{\acrlong{aaa}}
\label{sec:AAA}

\begin{figure} 
  \centering
  \begin{subfigure}{\textwidth}
    \centering
    \includegraphics[width=0.7\linewidth]{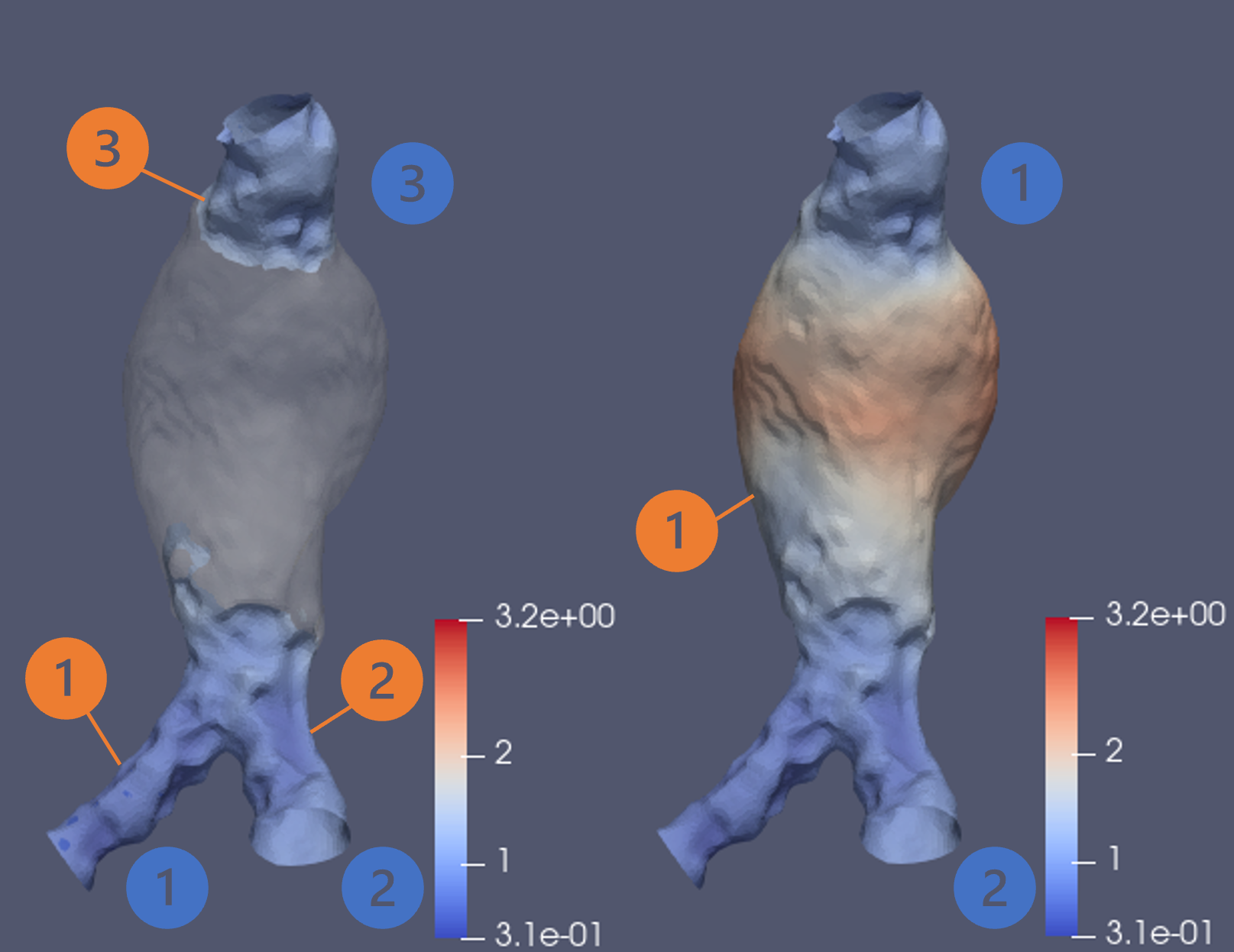}
    \caption{Two different steps of the sublevel set filtration of a patient affected by AAA: in the first step we clearly see the two path connected components - orange (1), which is also merged with (2), and orange (3) - which are separated by the AAA. At very high filtration values the AAA is added to the filtration and the path connected components merge. A similar phenomenon involves also 1-cycles: on the left, the loop going around the upper portion of the aorta cannot \virgolette{slide} down on the mesh and be equivalent to the loop going around the lower portion of the aorta - made by (1)+(2). This is instead possible on the right, causing the death of loop (3) at high values.}
    \label{fig:AAA_filtr}
  \end{subfigure}
  \begin{subfigure}{\textwidth}
    \centering
    \includegraphics[width=0.7\linewidth]{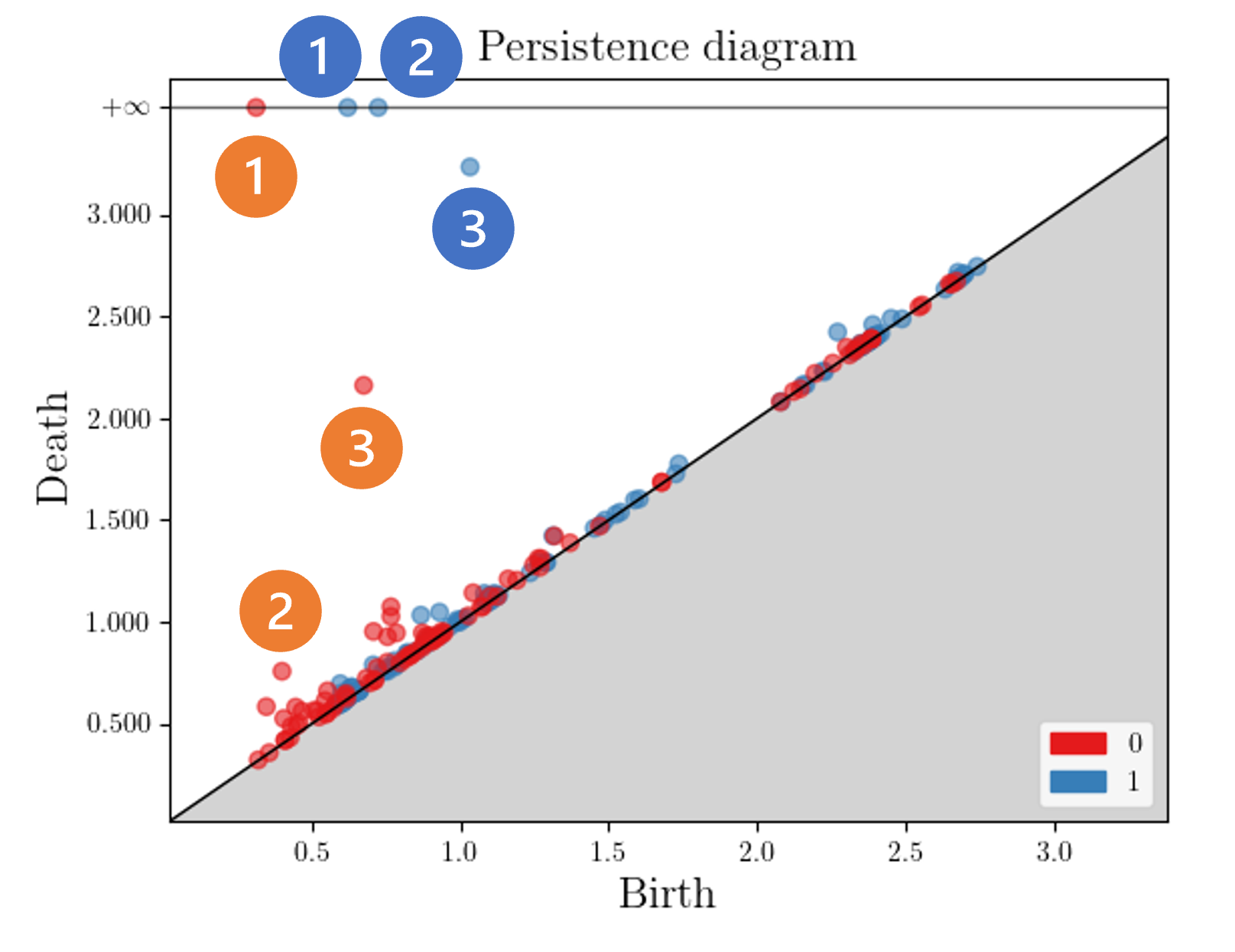}
    \caption{Persistence diagram of the filtration in \Cref{fig:AAA_filtr}. We have highlighted the high persistence path connected component and loop caused by the AAA splitting in two parts the aorta.}
    \label{fig:AAA_PD}
  \end{subfigure}

  \caption{The merging phenomena associated to \acrshort{aaa}.}
  \label{fig:AAApdgm}
\end{figure}

We now focus on \acrshort{aaa}s, which are obviously distinguished by large radial distances from the centerline and therefore by high values of birth and death coordinates of their associated points on the persistence diagram - see also \Cref{fig:AAApdgm}.   
But more can be said; 
in fact, persistence diagrams of patients affected by \acrshort{aaa} are characterized by the following persistence pairs:

\begin{enumerate}
    \item A notable and highly persistent class of $0$-cycles is usually present when patients are affected by an \acrshort{aaa}. These cycles are associated to the sudden increase of the distance from the centerline when moving form the aortic neck region to that part of the vessel where the \acrshort{aaa} occurs. In fact, the presence of the \acrshort{aaa} \virgolette{splits} the aorta in two parts, separated by the AAA - see \Cref{fig:AAApdgm}. Along the filtration, each part generates a connected component; those two components merge only when it is possible to have a path across the aneurysm; see \Cref{fig:AAApdgm}.

    \item Similarly, the \acrshort{aaa} generates also a notable highly persistent class of 1-cycles, associated with the tubular structure of the neck,  which doesn't appear in healthy patients. As already mentioned AAA splits the blood vessel in two parts - see \Cref{fig:AAApdgm}. Each part has a loop going around the aorta. Only when all the aneurysm appears in the filtration the loop around the blood vessel on one side of the aneurysm and the one on the other, can be merged with each other and become equivalent. One of them therefore dies and the other one persists as structural persistence pair. 
\end{enumerate}
    
    As a consequence, the corresponding point on the persistence diagram is easily recognizable since its birth is around the value $1.0$ and its death has the highest value among all the points with finite persistence. Again, all of this is clearly visible in \Cref{fig:AAApdgm}.

To summarize, two main features make the persistence diagram of a patient with \acrshort{aaa} easily distinguishable - see also \Cref{fig:AAApdgm}:

\begin{enumerate}
    \item The presence of persistence pairs borning and dying at high distance values. In fact, no healthy aorta has a region with a distance from the centerline higher than 1.4, while the regions of an aneurysmal aorta can exceed the value of 7.0.
    \item The presence of at least one $0$-cycle and one $1$-cycle classes with high persistence in the neck area (birth around $1$). 
\end{enumerate}

\subsubsection{Calcifications}
\begin{figure} 
  \centering
    \begin{subfigure}{0.4\textwidth}
    \centering
    \includegraphics[width=\textwidth]{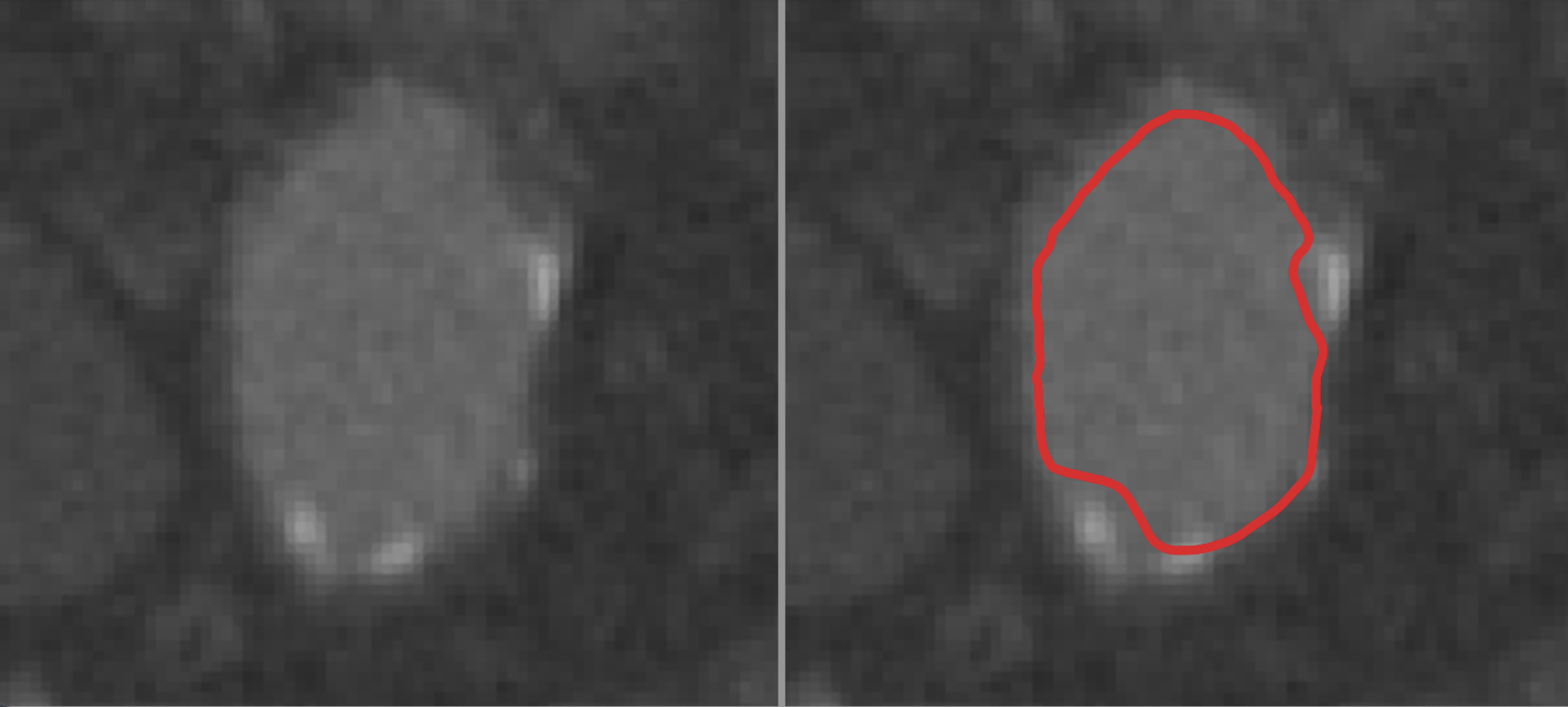}
    \caption{\acrshort{cta} slice of a non-aneurysmal aorta with presence of calcifications which creates bump in the lumen of the blood vessel.}
    \label{fig:calc_cta}
    \end{subfigure}
  \begin{subfigure}{0.55\textwidth}
    \centering
    \includegraphics[width=\textwidth]{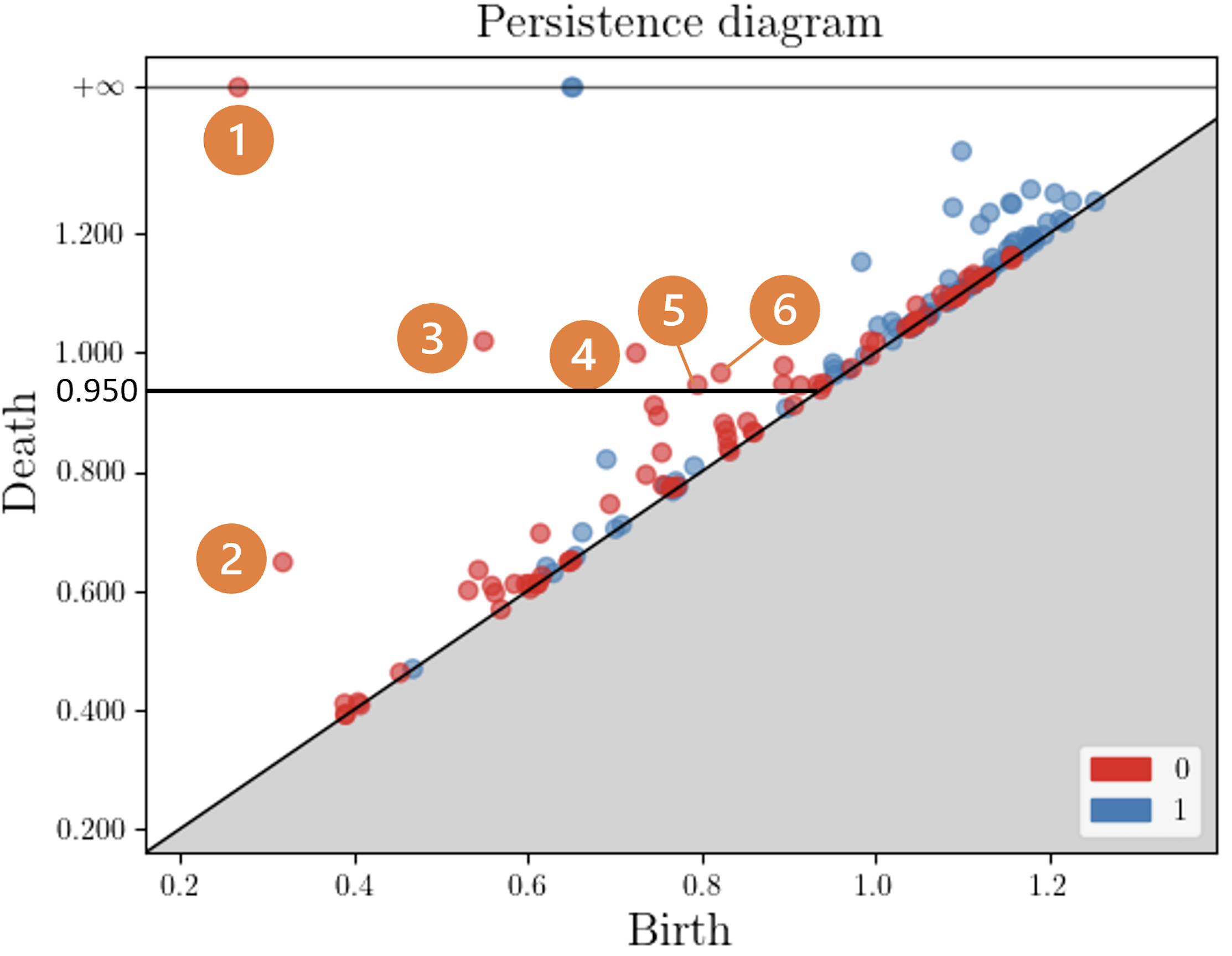}
    \caption{Persistence diagram of the filtration in \Cref{fig:calc_filtr}: we see a number of persistence pairs with medium persistence, being born in the main body of the aorta - i.e. with birth value higher than the aortic neck, as a consequence of the irregularities due to calcifications.}
    \label{fig:calc_PD}
  \end{subfigure}
  
  \centering
  \begin{subfigure}{\textwidth}
    \centering
    \includegraphics[width=1\textwidth]{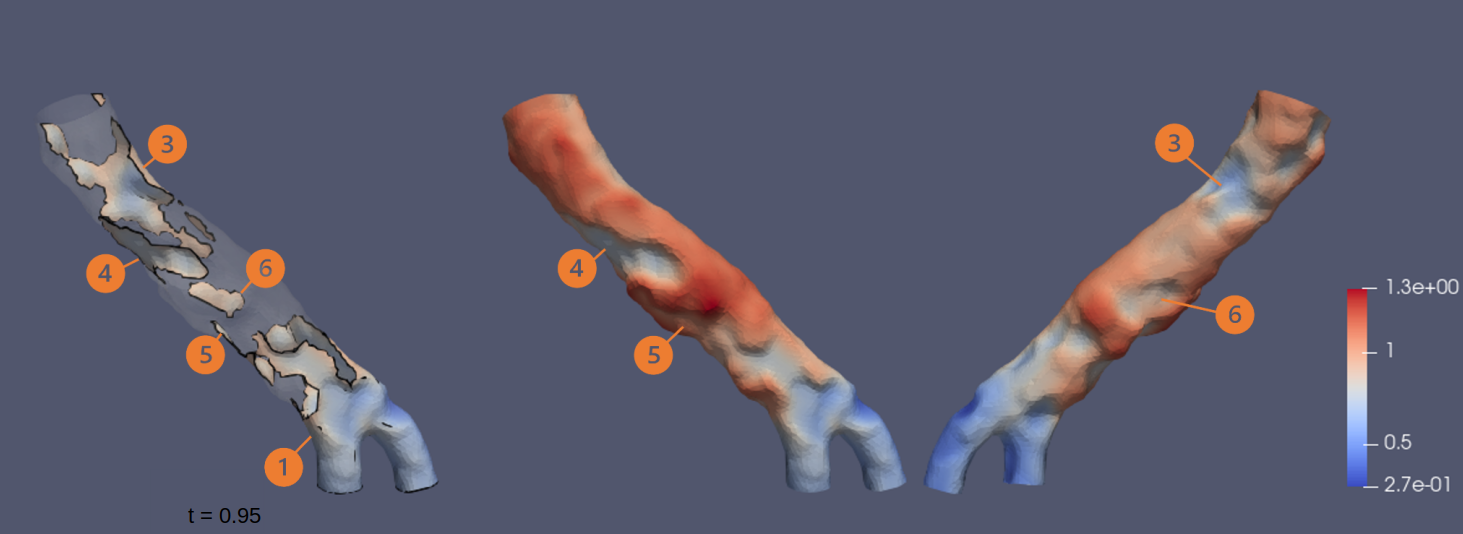}
    \caption{Different time steps of the sublevel set filtration of an aorta without AAA but with calcifications: calcifications create many medium-sized irregularities and bumps in the lumen which are picked up by the filtration, creating a number of path connected components arising and persisting for some time.}
    \label{fig:calc_filtr}
  \end{subfigure}

  \caption{Non-aneurysmal aorta with calcifications.}
  \label{fig:calcifications}
\end{figure}

    Calcifications are calcium deposits creating a local thickening of the aortic wall, on the internal side, resulting in indentations on the lumen's surface, when in contact with the calcified wall. Thus, they produce local minima in the filtering function, as showcased in \Cref{fig:calcifications}. These inward bumps are picked up as 0-cycles with medium-persistence in the persistent diagram; see \Cref{fig:calc_PD}.

\subsubsection{Thrombus}\label{sec:thrombus}
\begin{figure}
    \centering
    \includegraphics[width=\textwidth]{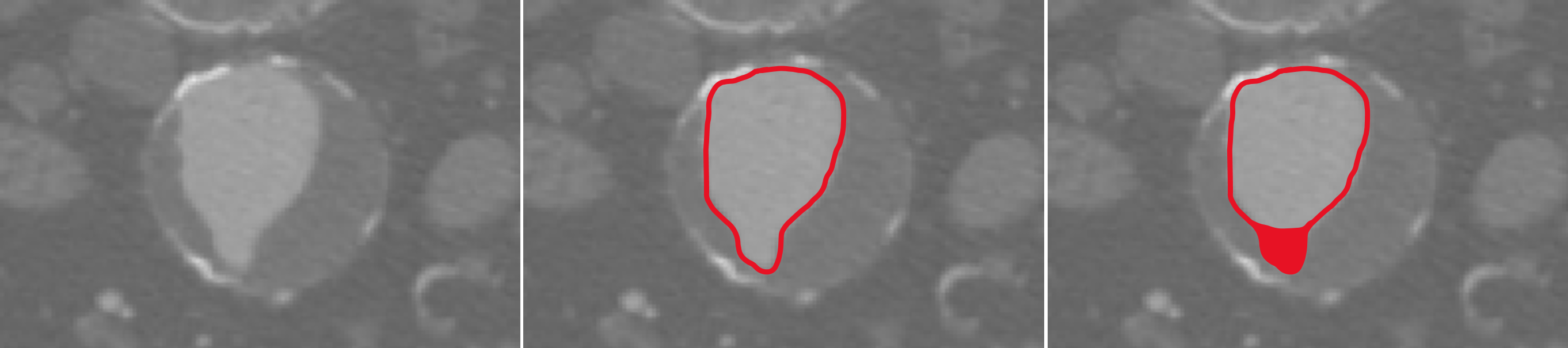}
    \caption{\acrshort{cta} slice of an \acrshort{aaa} with thrombus occluding the lumen.}
    \label{fig:thrombus}
    \end{figure}

    Lastly we consider blood vessels affected by a thrombus. A thrombus is often made by a non-homogeneous substance, which can flake off with the flow of blood creating wide ledges in the lumen (see \Cref{fig:thrombus}). It is most likely to be located in the area affected by \acrshort{aaa}, due to the concentration of metabolites associated with inflammation. Thus, the presence of a pronounced thrombus produces a series of local maxima and/or minima, which induce 1 and 0 cycles in the persistence diagrams with relatively high birth coordinates and moderate persistence, see also \Cref{fig:PD_complete_example}. 
    
\subsection{A detailed Example}\label{sec:complete_example}

We now show an example of a concerning patient affected by \acrshort{aaa}, iliac aneurysm and thrombus.

\begin{figure} 
    \centering
    \includegraphics[width=0.8\textwidth]{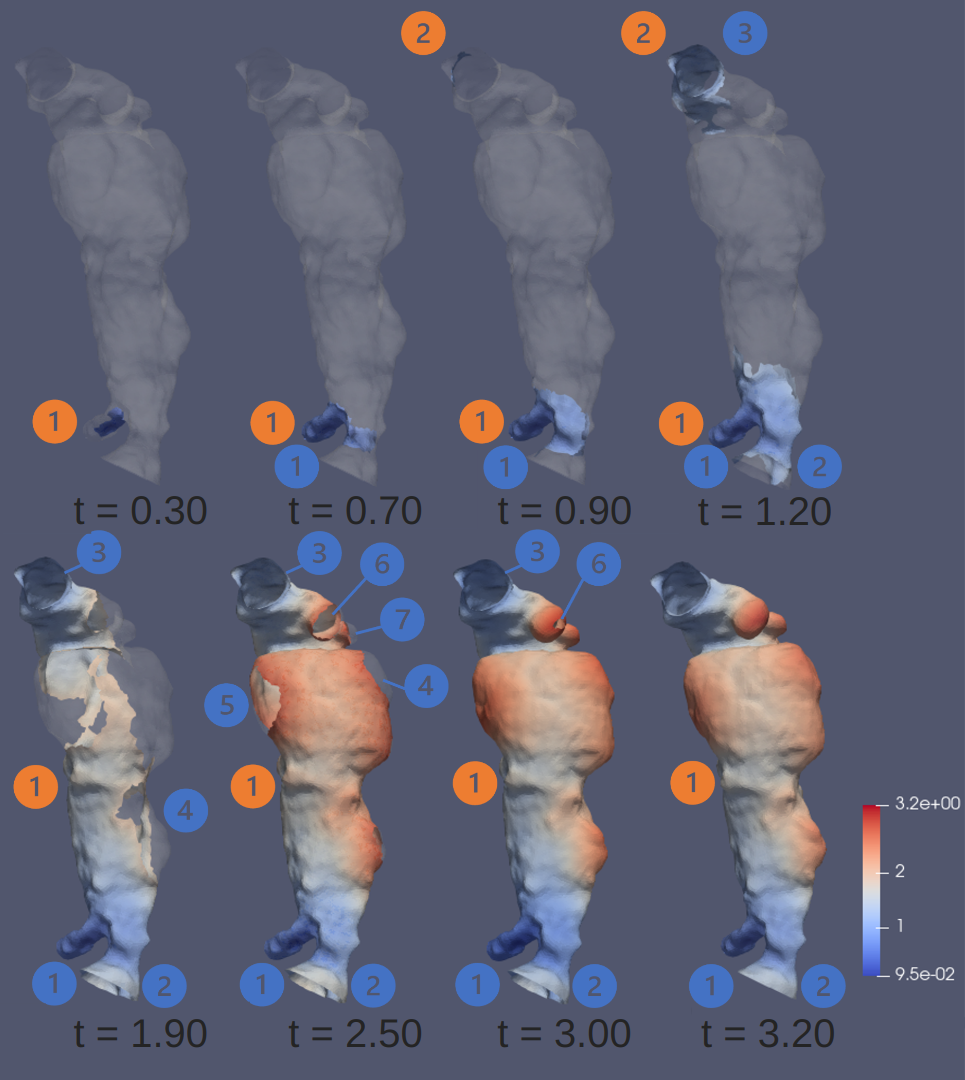}
    \caption{A patient affected by \acrshort{aaa}, iliac aneurysm and thrombus. The filtration function values are reported in black on the bottom left corner of the aorta. The sublevel sets are represented by the higlighted portions of the mesh. The numbered features identify $0$-dimensional (red) and $1$-dimensional cycles associated to points in the diagram appearing in \Cref{fig:PD_complete_example}.}
    \label{fig:example}
    \end{figure}

\begin{figure} 
    \centering
    \includegraphics[width=\textwidth]{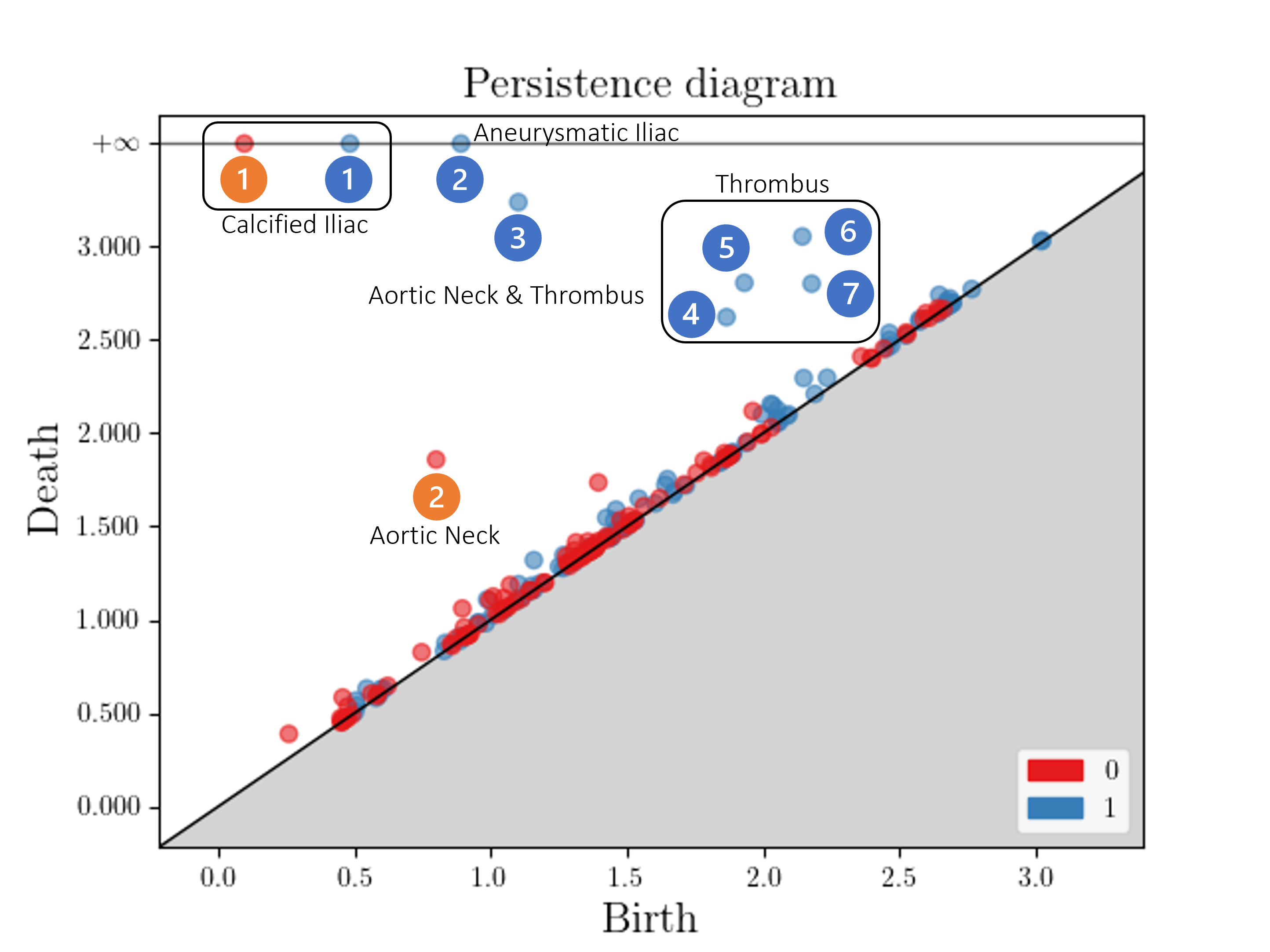}
    \caption{The persistence Diagram associated with the mesh shown in \Cref{fig:example}. Features labelled as (1) both in dimension 0 and 1 are the structural pairs associated to the iliac which does not present an aneurysm. In particular, path connected component (1) is born very early along the filtration and that is because of some inwards bumps due to some calcifications. Persistence pairs (2) - both in dimension 0 and 1 -, instead, are born very late, almost at the filtration value of the neck. This is a clear consequence of the iliac aneurysm. The 1-cycles' class (3) is generated because of the AAA splitting the blood vessel in two parts. An in fact it dies at the latest value of the filtration. Then we have a group of 1-cycles - from (4) to (7) - with medium persistence, which reflect the irregularities in the lumen caused by some heterogeneous thrombus.}
    \label{fig:PD_complete_example}
    \end{figure}

In \Cref{fig:example} we report the main steps of the sublevel set filtration of the selected patient. While in \Cref{fig:PD_complete_example} the associated persistence diagram is shown.

\begin{itemize}
    \item At value $0.3$ a 0-cycle of the first iliac appears.
    \item At value $0.7$ a $1$-cycle, around that iliac, appears. Note that the 0-cycles expands onto the other iliac without creating a new path connected component. There is in fact an aneurysm on the second iliac preventing the tapering of its extremity.
    \item At value $0.9$ a novel 0-cycle appears on the aortic neck. The AAA separates this cycle from the one appeared in the beginning.
    \item At value $1.20$ a $1$-cycle going around the blood vessel appears. 
    \item As the distance form the centerline increases - that is, the value of the radial filtration function increases -, the initial path connected component expands enclosing gradually also the AAA, connecting to and killing the path connected component and the loop associated to the neck. On the way, several classes of $1$-cycles are created with high birth times; they are associated to the local maxima of the filtration function created in the aneurysmatic portion of the aorta by the thrombus. 
\end{itemize}

\section{Population Analyses}\label{sec:analyses}

The previous section discussed how persistent homology translates some relevant aortic features into persistence pairs shown as points of a persistence diagram. We now explore the behaviour of these topological summaries at the population level, to understand which kind of between patients variability can be captured with persistence diagrams.

We present two paradigmatic problems, both characterized by their simplicity: clustering and discrimination or, with different words, unsupervised and supervised classification of persistence diagrams. 
In the initial clustering exercise, we look for a natural stratification of persistence diagrams embedded in a metric space endowed with a Wasserstein metric.
Secondly, we exploit the analysis of \Cref{sec:read_PDs} to introduce a supervised classification pipeline aimed at the construction of classifiers identifying patients with: (1) AAA, (2) calcifications, (3) thrombus, (4) iliac aneurysm.

\subsection{Clustering}\label{sec:clustering}

Throughout the manuscript we argued that irregularities in the aortas are represented by inward and outward bumps, representing pathologies of different nature, which are captured separately by 0-cycles or 1-cycles represented as persistence pairs in a persistence diagram. 

Since both the persistent diagrams for classes of 0-cycles and 1-cycles are relevant to capture shape differences between aortas, we introduce the following family of metrics between patients $P_i$ and $P_j$:
\[
d_p^p(P_i,P_j) = \lambda \cdot W_p^p(D_i^0,D_j^0) + (1-\lambda)\cdot W_p^p(D_i^1,D_j^1),
\]
where $p\geq 1,$ \(\lambda \in [0,1]\) and, for \(k=0,1,\) the persistence diagrams $D_i^k, D_j^k $ are those for the \(k\)-cycles of patient $i$ and \(j\) respectively; \(W_p\) is the p-Wasserstein distance introduced in \eqref{p_Wasserstein_dist}.
Note that, by setting $p=\infty,$ one obtains the weighted average of the bottleneck distances between the diagrams.
Although different mixing weights are allowed, in the following  we set \(\lambda=0.5\) for simplicity. Tailoring the choice of \(\lambda\)  could benefit the analysis of specific classification problems.  

\begin{figure}
\begin{subfigure}{0.4\textwidth}
    \centering  
    \includegraphics[width = \textwidth]{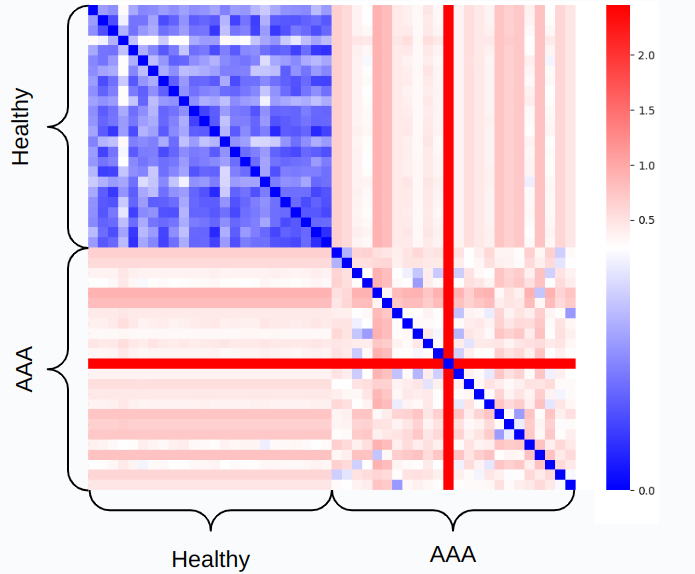}
    \caption{Matrix of pairwise p-Wasserstein distances between persistence diagrams, with \((p=\infty)\). Patients are ordered so that healthy patients come before non-healthy ones.}
    \label{fig:clustering_B}
\end{subfigure}
\begin{subfigure}{0.4\textwidth}
    \centering  
    \includegraphics[width = \textwidth]{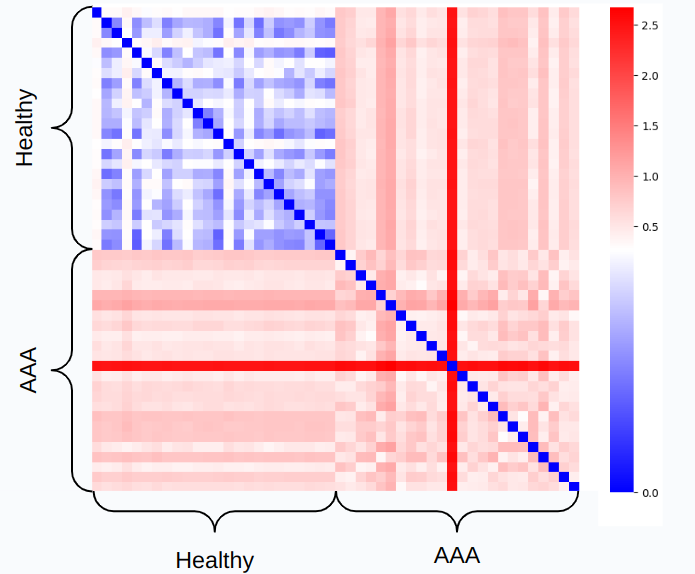}
\caption{Matrix of pairwise p-Wasserstein distances between persistence diagrams, with \((p=2)\). Patients are ordered so that healthy patients come before the ones affected by \acrshort{aaa}.}
\label{fig:clustering_W}
\end{subfigure}

\begin{subfigure}{0.48\textwidth}
    \centering  
    \includegraphics[width = \textwidth]{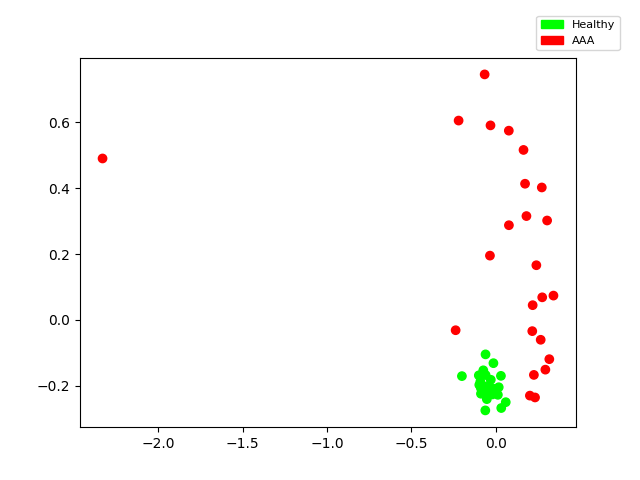}
\caption{Two dimensional \acrshort{mds} representation of the matrix in \Cref{fig:clustering_B}.}
\label{fig:clustering_B_MDS}
\end{subfigure}
\begin{subfigure}{0.48\textwidth}
    \centering  
    \includegraphics[width = \textwidth]{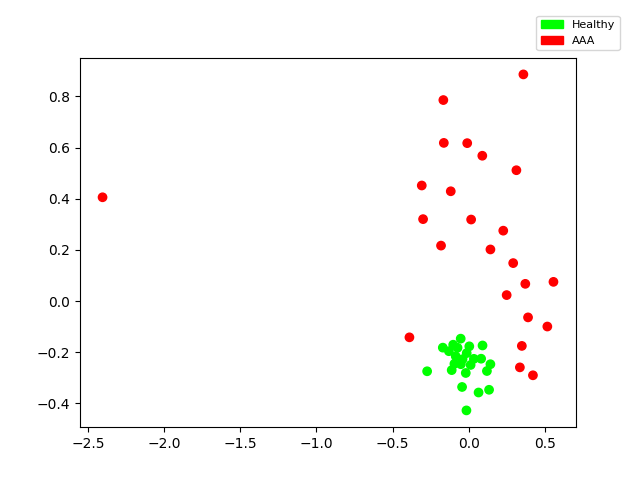}
    \caption{Two dimensional \acrshort{mds} representation of the matrix in \Cref{fig:clustering_W}.}
    \label{fig:clustering_W_MDS}
\end{subfigure}
\caption{Pairwise distance matrices and low-dimensional embeddings for the cohort of patients described in \Cref{sec:clustering}.}
\label{fig:clustering}
\end{figure}

\Cref{fig:clustering} depicts the matrices of pairwise distances between patients in the cohort of our study, when $p=\infty$ and $p=2$  respectively. Statistical units have been ordered so that healthy patients come before those with an AAA: indeed, $H01,\ldots,H24$  index the healthy patients whithout AAA while $A01,\ldots,A24$ indicate the patients with AAA. By visual inspection, it is clear, both in \Cref{fig:clustering_W} and \Cref{fig:clustering_B}, that this grouping is very well captured by both distance matrices. In fact, visualizing a low-dimensional representations, obtained by MultiDimensional Scaling (MDS), of the metric spaces embedding the persistence diagrams -- see \Cref{fig:clustering_W_MDS} and \Cref{fig:clustering_B_MDS} -- we immediately notice how patients without AAA are clustered together, while the presence of an AAA, by  introducing high persistence points in the persistence diagram, generates a larger variability in the MDS representation of the persistence diagrams of diseased patients - see \Cref{sec:AAA}. 

\begin{figure}
    \centering  
\begin{subfigure}{0.49\textwidth}
    \centering  
    \includegraphics[width = \textwidth]{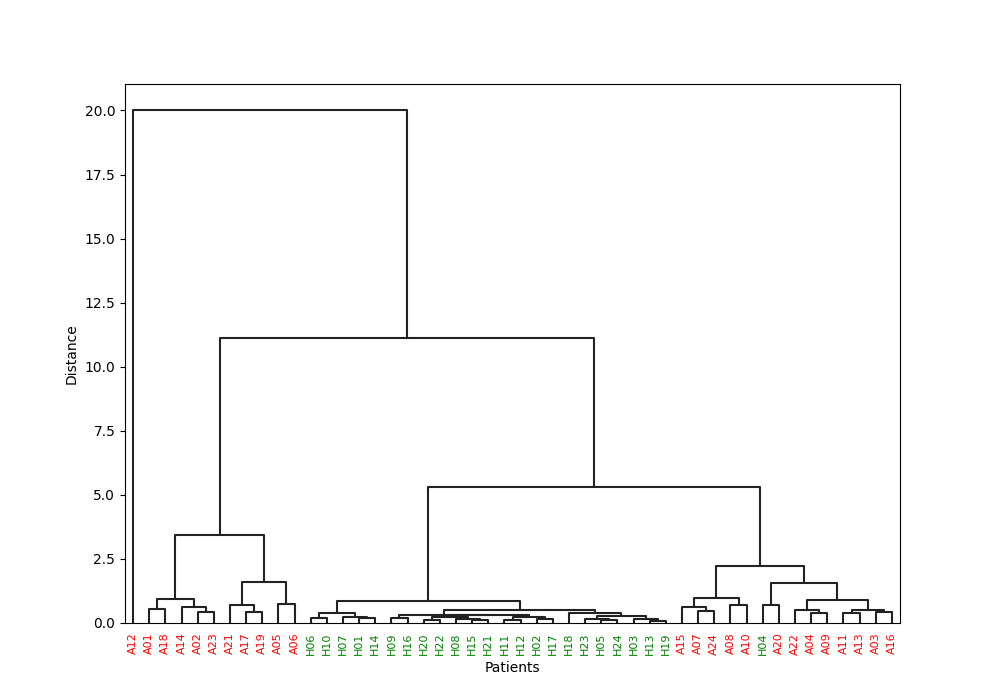}
\caption{Hierarchical clustering dendrogram with the previously defined Bottleneck distance and Ward linkage.}
\label{fig:dendro_B}
\end{subfigure}  
\centering  
\begin{subfigure}{0.49\textwidth}
    \centering  
    \includegraphics[width = \textwidth]{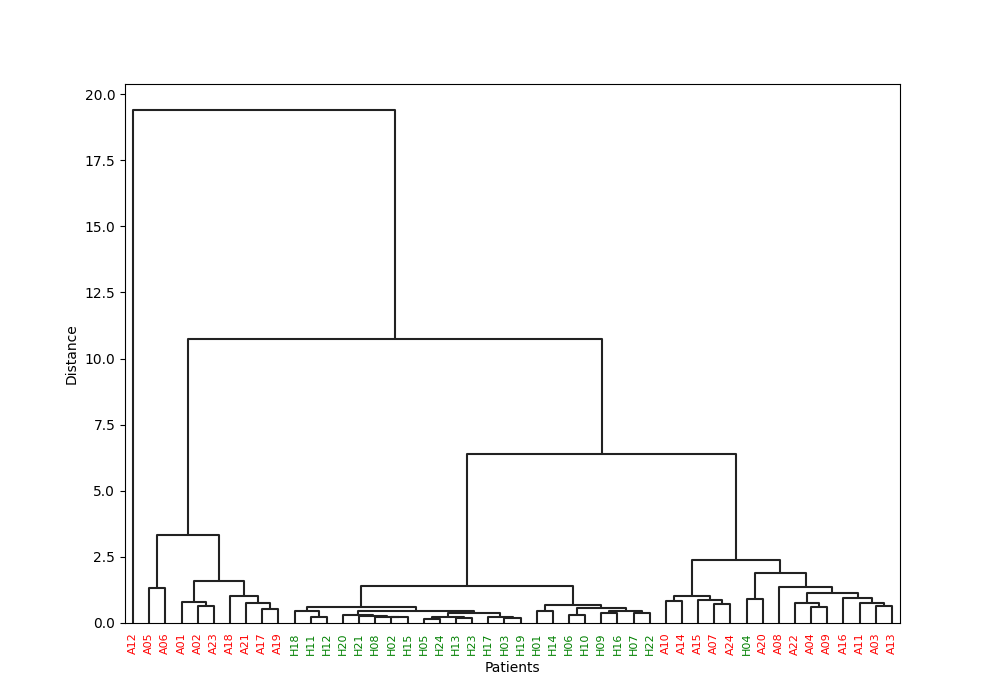}
\caption{Hierarchical clustering dendrogram with the previously defined 2-Wasserstein distance and Ward linkage.} 
\label{fig:dendro_W}
\end{subfigure}

\caption{Hierarchical clustering dendrograms related to the clustering problem presented in \Cref{sec:clustering}. Patients in red are affected by AAA, while healthy 
ones are drawn in green.}
\label{fig:dendros}
\end{figure}

This grouping can indeed be captured by a clustering algorithm. For instance we report the dendrograms relative to agglomerative hierarchical clustering run with Ward linkage, in \Cref{fig:dendro_W} and \Cref{fig:dendro_B}. Leaves are coloured as in \Cref{fig:clustering_W_MDS} and \Cref{fig:clustering_B_MDS}.

\subsection{Supervised classification - AAA}\label{sec:AAA_class}
In \Cref{sec:clustering} we reported an unsupervised algorithmic pipeline which generates a very clear clusterization of the topological representations of the patients in our study. We now setup a supervised classification algorithm aimed at discriminating patients with an AAA from the others.  

\begin{figure}
    \centering  
\begin{subfigure}{0.45\textwidth}
    \centering  
    \includegraphics[width = \textwidth]{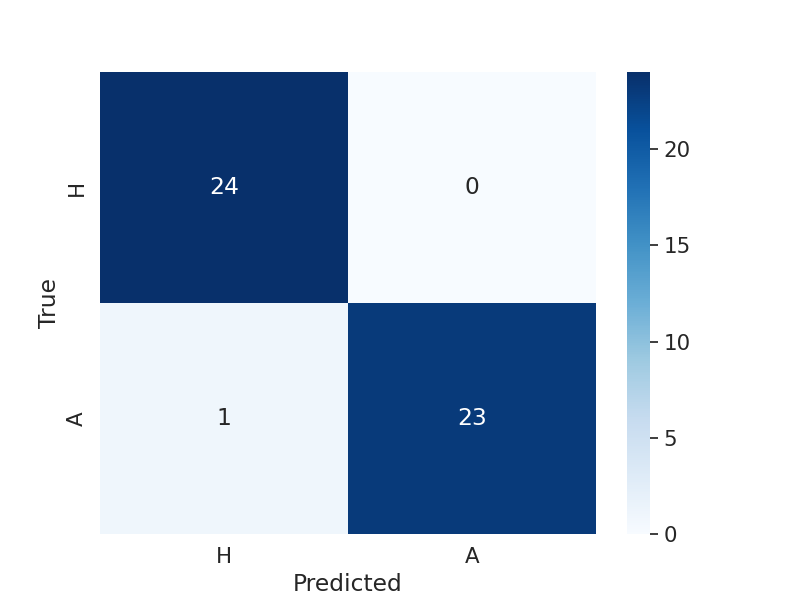}
\caption{Leave-one-out confusion matrix for the classification case study presented in \Cref{sec:AAA_class}, starting from the matrix obtained with the bottleneck distance.}
\label{}
\end{subfigure}
\centering  
\begin{subfigure}{0.45\textwidth}
    \centering  
    \includegraphics[width = \textwidth]{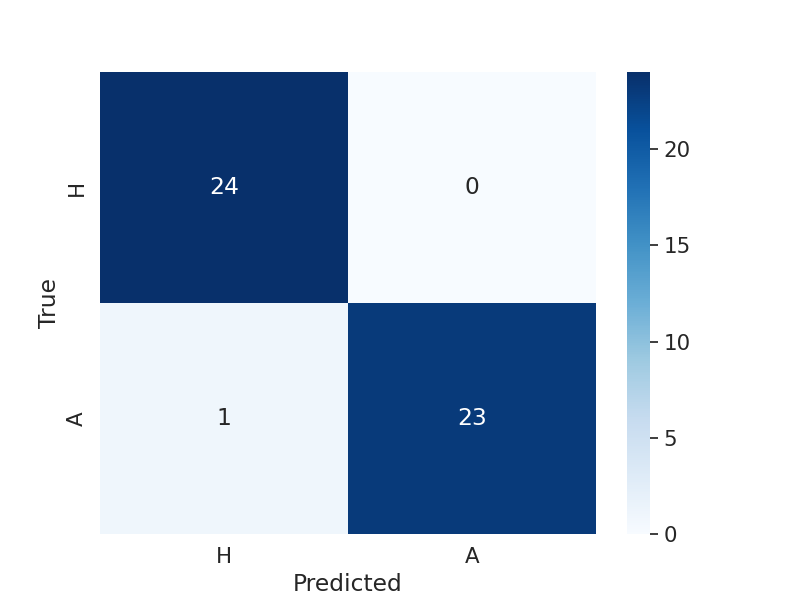}
\caption{Leave-one-out confusion matrix for the classification case study presented in \Cref{sec:AAA_class}, starting from the matrix obtained with the 2-Wasserstein.}
\label{}
\end{subfigure}

\caption{Confusion matrices for the classification case study presented in \Cref{sec:AAA_class}. Label A is for patients with AAA, H for the "healthy" ones.}
\label{fig:knn_confusion}
\end{figure}

The dendrograms in \Cref{fig:dendros} justify the use of a very simple algorithm like k-nearest neighbors (KNN). We report the leave-one-out (l1out) confusion matrices in \Cref{fig:knn_confusion} for $k=5$. We point out that for any choice of $k\in \{1,\ldots,5\}$ we get $97\%$ accuracy, with the only missclassified patient being always the same one, visibly close to the healthy patients also in \Cref{fig:clustering}. This patient has been checked by the collaborating clinicians and it has been regarded as having a relatively small AAA.

\subsection{Supervised classification - Calcifications}
\label{sec:calcifications}

We now want to discriminate patients with calcifications from the others. To do so, we rely on \Cref{sec:calcifications}: therein we showed that calcifications appear as mid-persistence features in the zero dimensional persistence diagram, being local minima of the radial filtering function. 

\begin{figure}
    \centering  
    \includegraphics[width = \textwidth]{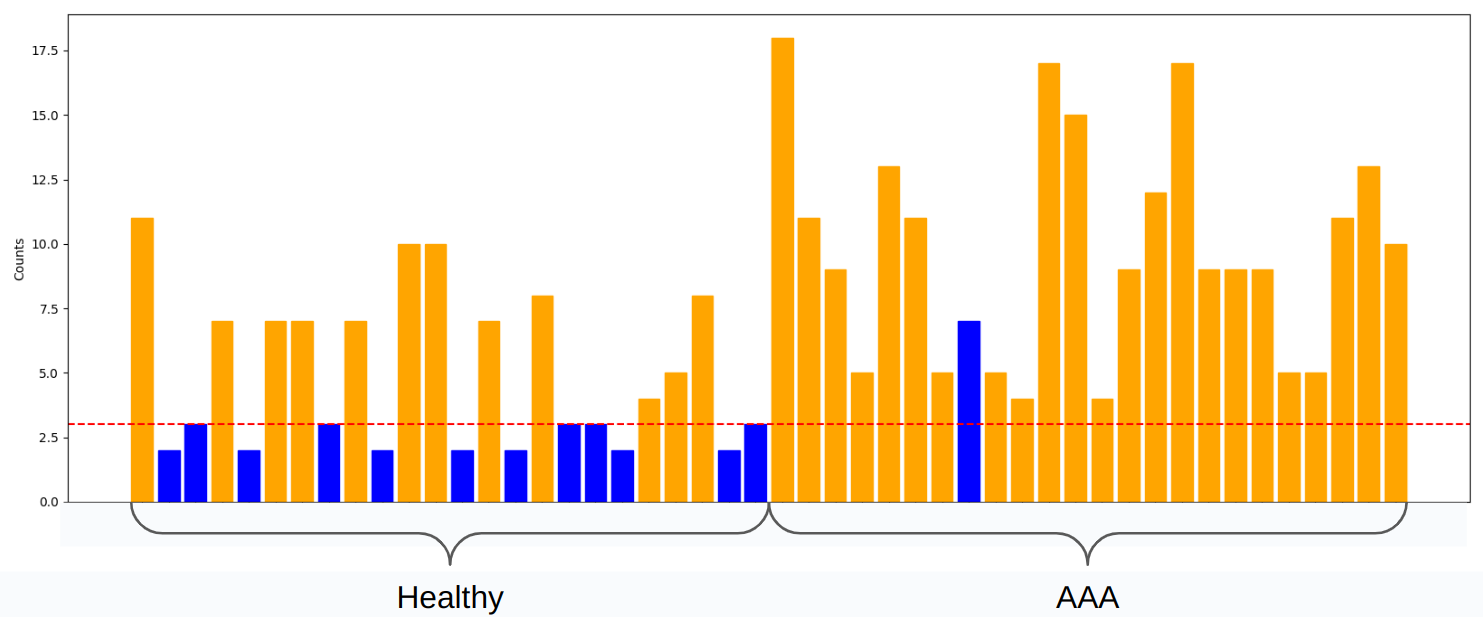}
\caption{Plot related to \Cref{sec:calcifications}, representing the counts of the persistence pairs in dimension $0$ with persistence greater than $0.1$ to recognize patients with calcifications. Patients which present calcifications are plotted in orange, the others in blue. There is clearly correlation between the presence of calcifications and irregularities of the lumen captured by 0 dimensional homology. And we highlight that this holds true also for healthy patients, which is arguably the more interesting situation, as most of the patients affected by AAA present also calcifications.
Patients are labelled and ordered as in previous figures.}
\label{fig:calcifications_counts}
\end{figure}

Thus, for all patients, we count the number of 0 dimensional persistence pairs with persistence greater than or equal to a threshold and less than $\infty$. Patients with or without calcifications have been identified and labelled by looking at their TAC, with the help of a collaborating clinician; the  threshold \(\tau_p\) for persistence has been set to $0.1$, a value coherent with \Cref{sec:normalization_effect}.
Results are shown in \Cref{fig:calcifications_counts}: all patients whose count of filtered persistence pairs exceeds the threshold \(\tau_c\) of 3 are classified as affected by calcification. 
Both thresholds \(\tau_p\) and \(\tau_c\) have been selected to maximize leave-one-out accuracy; the values \(\tau_p=0.1\) and \(\tau_c=3\) result in only one leave-one-out missclassified patient. Maximization has been obtained through grid search; for \(\tau_p\) the grid spans a range from 0.005 to 0.5, with increments of 0.005, while for \(\tau_c\) it considers the natural numbers from 1 up to 20. 

Note that, as shown by \Cref{fig:calcifications_counts}, identifying patients with calcifications is more of a problem for those without an AAA - the \(S\) patients in the left half of \Cref{fig:calcifications_counts} - since almost all patients with an AAA show also the presence of calcifications.

\subsection{Supervised classification - Iliac Aneurysm}
\label{sec:iliac_aneu_class}

\begin{figure}
\begin{subfigure}{0.59\textwidth}
    \centering  
    \includegraphics[width = \textwidth]{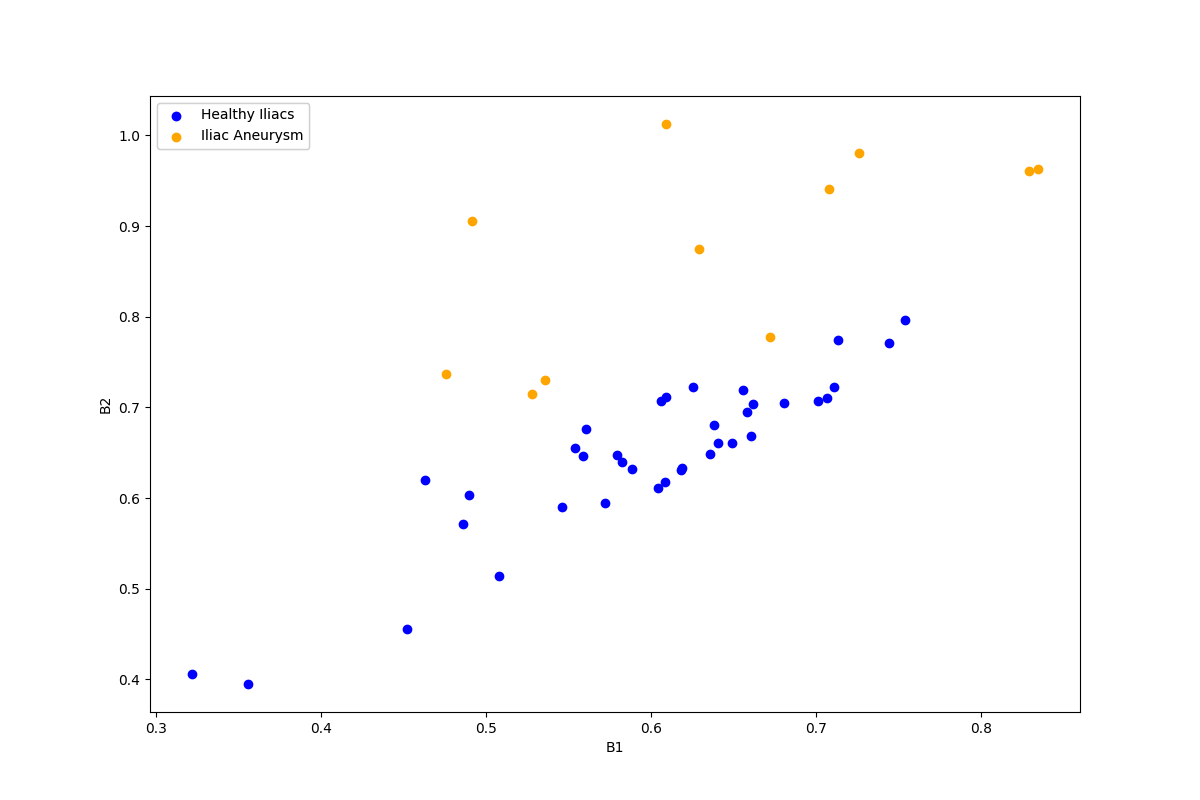}
\caption{Scatterplot of the variables $B_1$ and $B_2$ used for classifying patients with iliac aneurysm.}
\label{fig:lda_iliacs}
\end{subfigure}
\begin{subfigure}{0.4\textwidth}
    \centering  
    \includegraphics[width = \textwidth]{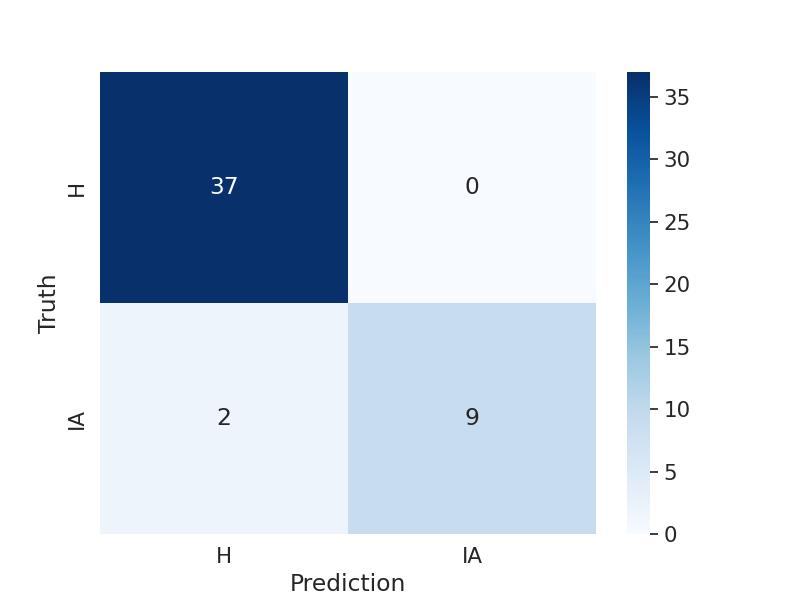}
\caption{L1out confusion matrix for Iliac Aneurysm classification. Label AI is for patients with iliac aneurysm, H for the others.}
\label{fig:iliacs_confusion}
\end{subfigure}

\caption{Iliac Aneurysm classification, \Cref{sec:iliac_aneu_class}.}
\label{fig:iliac_aneurysm}
\end{figure}

We now deal with iliac aneurysms. As discussed in \Cref{sec:iliacs},
we expect an aneurysm in the iliac area to show up in the persistence diagram in two different ways: 1) the absence of a class of 0-cycles born on the affected iliac, separately from the one encompassing the healthy one; 2) one or two classes of 1-cycles with infinite persistence, born at high times. Since the absence of the second class of 0-cycles is redundant with respect to the information of the birth of the second class of 1-cycles, we just consider as features for this classification problem the birth coordinates of the two classes of $1$-cycles with infinite persistence, ordered according to their birth. We call those two variables $B_1$ and $B_2$.  The difference between patients with or without an iliac aneurysm, it's evident in \Cref{fig:lda_iliacs}: large values of birth for both 1-cycles are associated to the presence of an iliac aneurysm. In particular the second coordinate, i.e. the largest birth, is the most discriminanting factor. This is coherent with the fact that having one iliac aneurysm increases the birth coordinate of the second class of $1$-cycles with infinite persistence; while two aneurysms increase both coordinates. Thus the second feature is always altered by the presence of an aneurysm at the iliacs.

We fit a linear discriminant analysis model; by leave-one-out, this predicts correctly all patients but two. The confusion matrix is reported in \Cref{fig:iliacs_confusion}.
Both misclassified patients are affected by iliac aneurysms, but classified as healthy by our pipeline: their aneurysms are indeed not severe, according to the collaborating clinicians.

\subsection{Supervised classification - Thrombus}\label{sec:class_thrombi}

Lastly, we turn to the problem of detecting thrombi, which have been discussed in \Cref{sec:thrombus}. We have argued that thrombi are characterized by irregularities in the lumen, most likely in the part of the blood vessel which is affected by AAA.
Depending on the homogeneity of the obstruction, these irregularities might generate local maxima of the radial filtering function.

We approach thrombus classification by filtering out points in the persistence diagram and retaining only 0 and 1 dimensional persistence pairs with finite persistence and high birth coordinates.

\begin{figure}
\includegraphics[width=\textwidth]{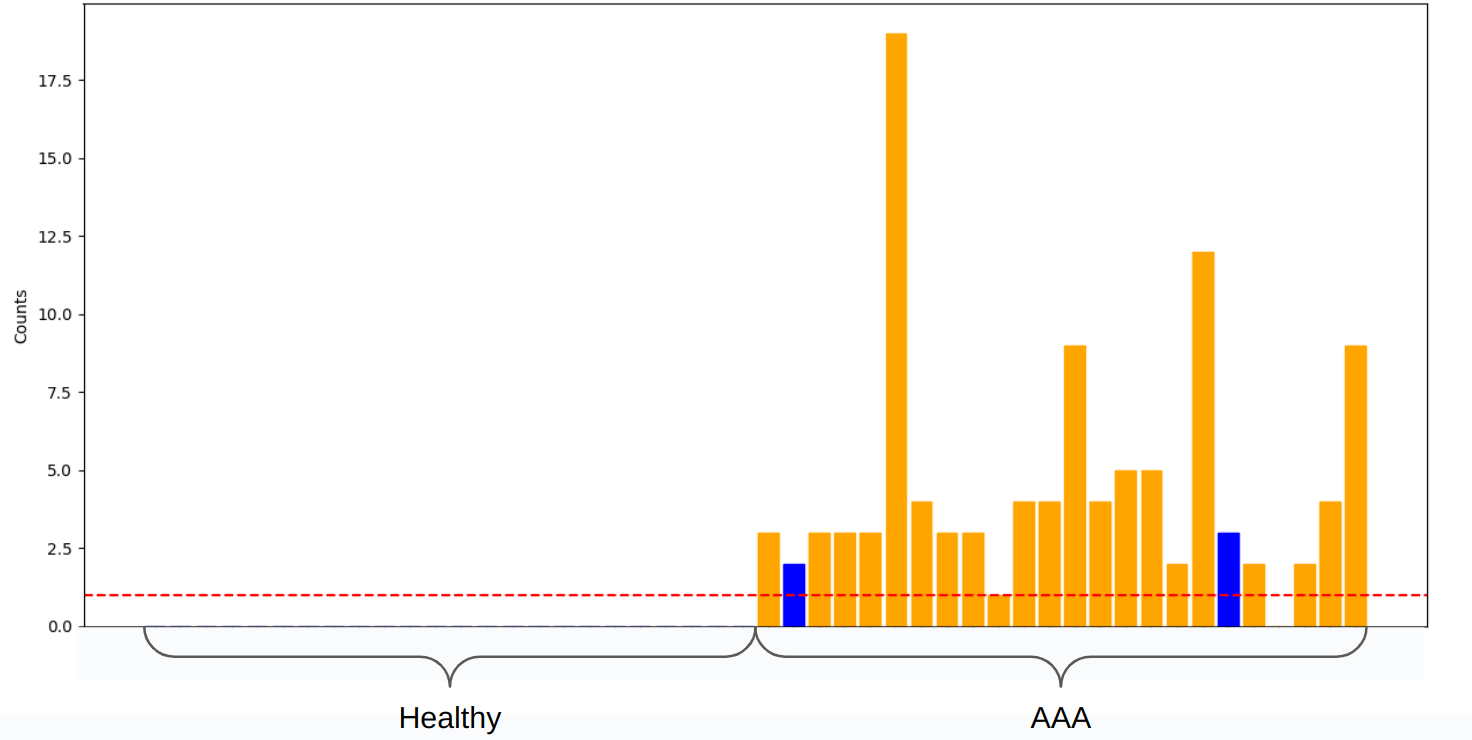}
\caption{Plot related to \Cref{sec:class_thrombi}, representing the counts of persistence pairs with medium persistence associated to irregularities of the lumen, trying to exclude bumps located on the iliacs and on the aortic neck. Patients which present thrombi are plotted in orange, the others in blue. Again we can see correlation between the presence of thrombi and irregularities of the lumen captured by the selected points in 0 dimensional homology.}
\label{fig:thrombi}
\end{figure}

\begin{figure}
\centering
\begin{subfigure}{0.49\textwidth}
\includegraphics[width=\textwidth]{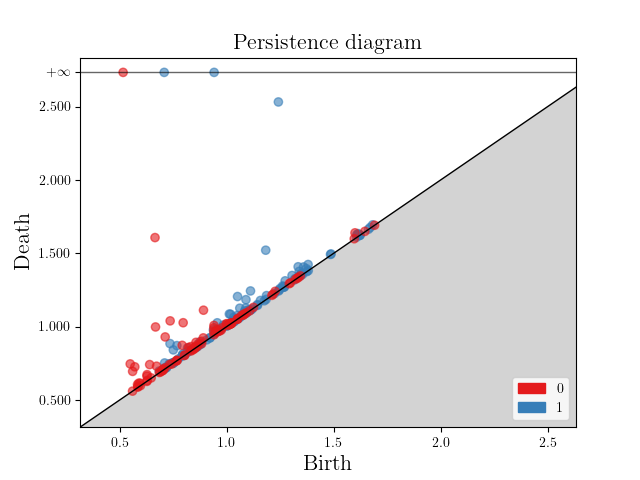}
\caption{Patient A20 missclassified by the analysis in \Cref{sec:class_thrombi}. }
\label{fig:PD_M20}
\end{subfigure}
\centering
\begin{subfigure}{0.49\textwidth}
\includegraphics[width=\textwidth]{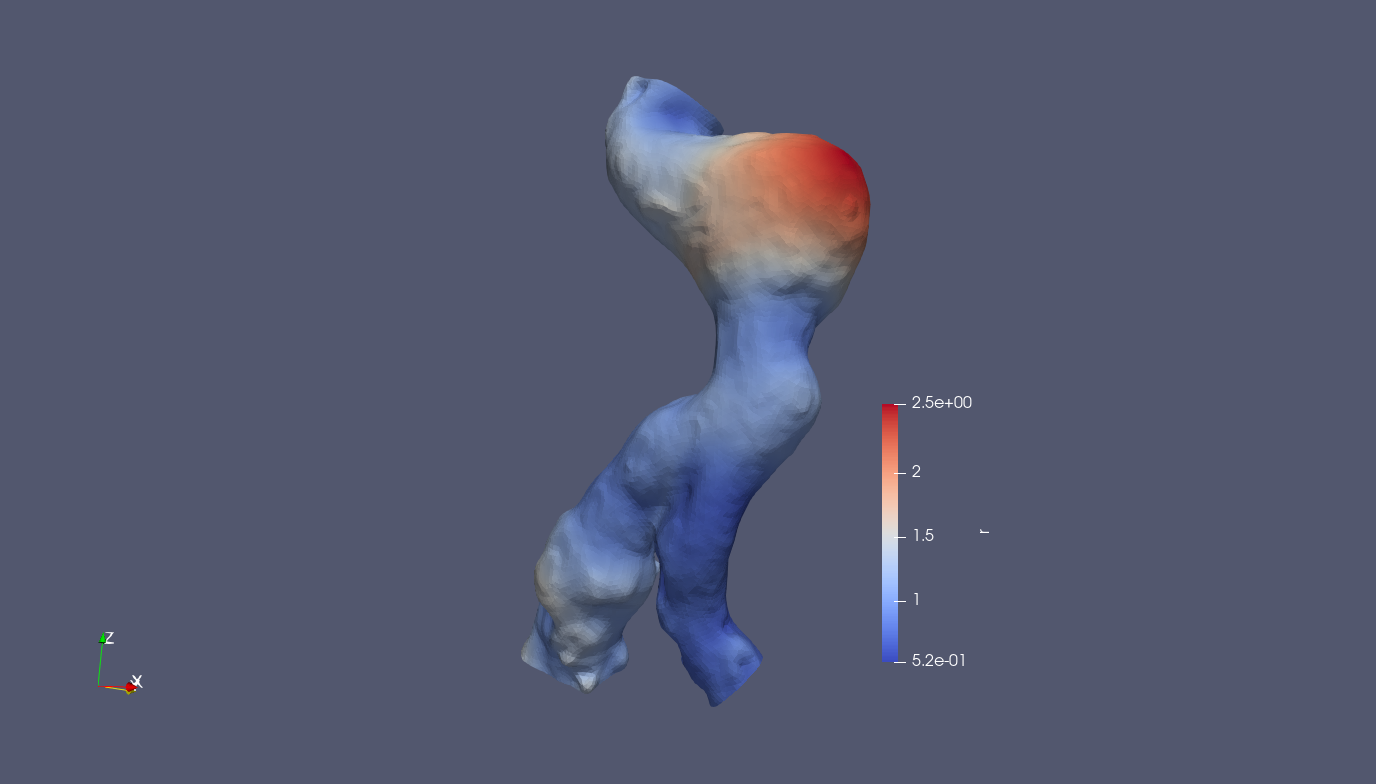}
\caption{The patient A20 missclassified by the analysis in \Cref{sec:class_thrombi}. }
\label{}
\end{subfigure}

\centering
\begin{subfigure}{0.49\textwidth}
\includegraphics[width=\textwidth]{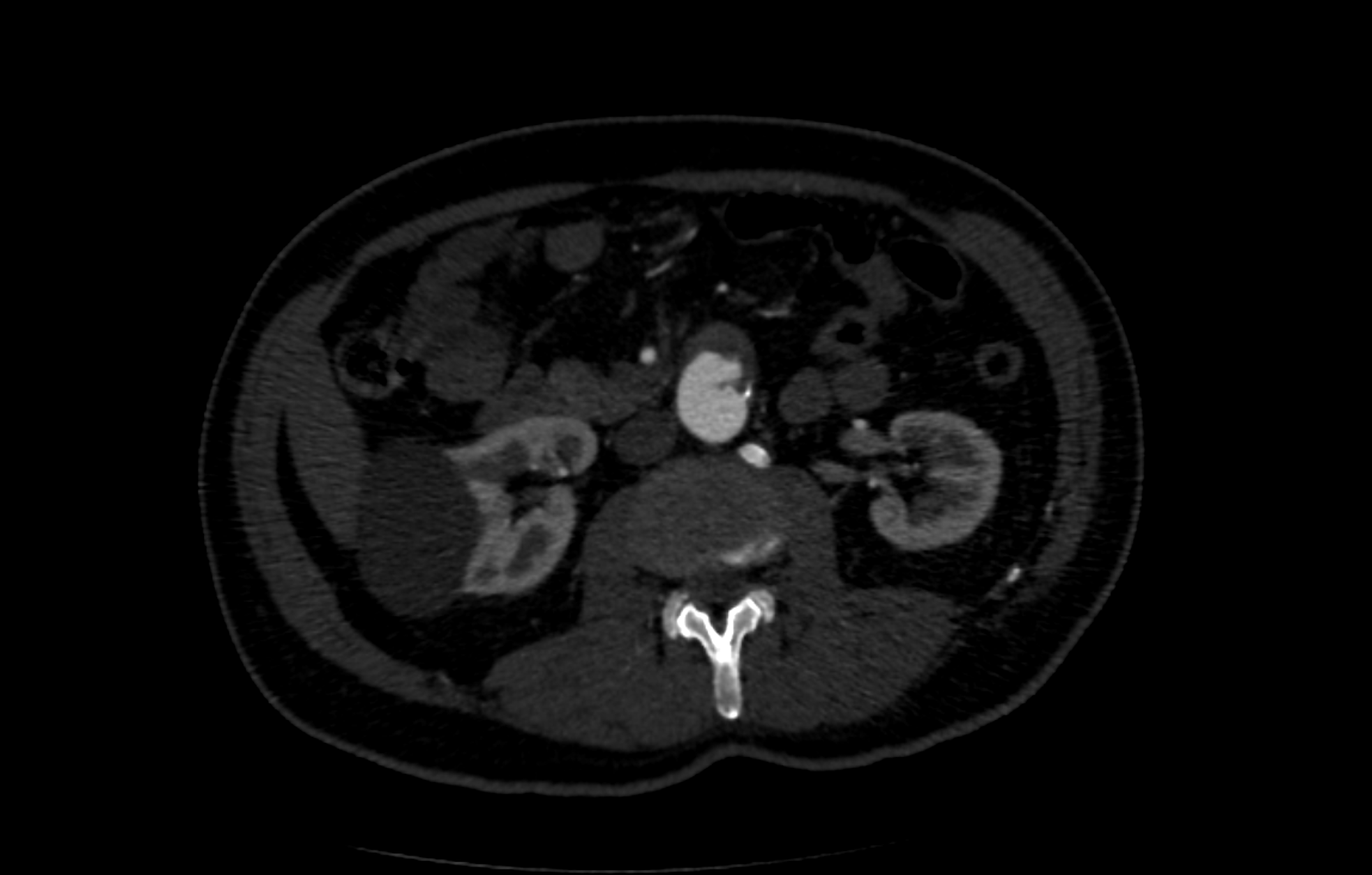}
\caption{Patient A10 missclassified by the analysis in \Cref{sec:class_thrombi}. }
\label{}
\end{subfigure}
\centering
\begin{subfigure}{0.49\textwidth}
\includegraphics[width=\textwidth]{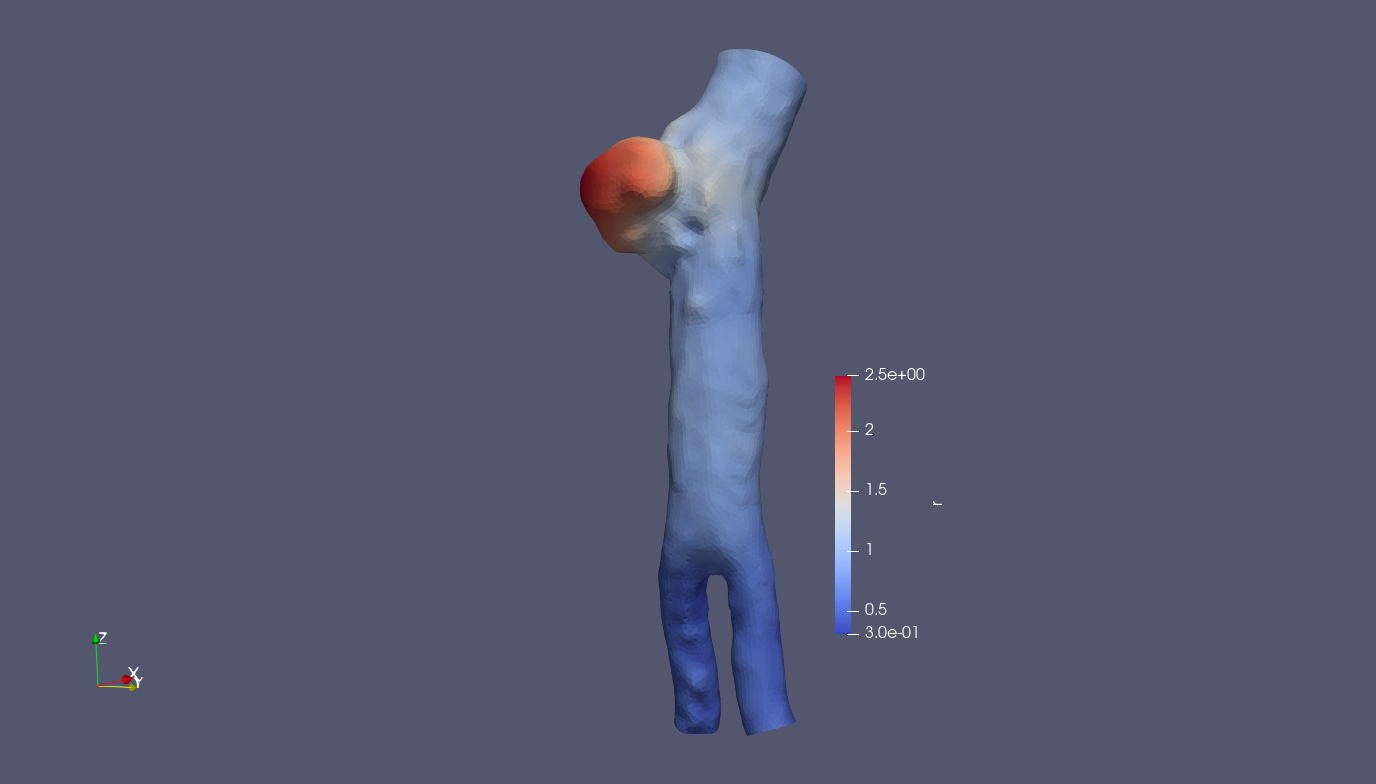}
\caption{Patient A10 missclassified by the analysis in \Cref{sec:class_thrombi}. }
\label{fig:M10}
\end{subfigure}
\caption{Patients missclassified by the analysis in \Cref{sec:class_thrombi}.}
\label{fig:missclass_thrombi}
\end{figure}

In particular, for each diagram, we retain 0-cycles and 1-cycles with birth coordinate larger than or equal to $1.1$ - so that we know we are not on the iliacs and the neck - and persistence larger then or equal to $0.1$ - to filter out smaller and noisy bumps. Disregarding the label declaring if they are 0-cycles or 1-cycles, the number of cycles retained for each patient is shown in \Cref{fig:thrombi}; note that all healthy patients do not present any thrombi. Thus, we can label as affected by thrombus all patients with at least $2$ retained classes of cycles, obtaining four missclassified patients: A10, A19, A20, A21. In \Cref{fig:missclass_thrombi} we report data about two of these patients, to understand the roots of this classification errors. For instance,
patient A20 is without thrombi but is classified as a having one. 
We can see in \Cref{fig:PD_M20} some high persistence features in dimension $1$, which are induced by a very irregular lumen, with multiple aneurysm-like dilations. But no thrombus. 
Instead patient A10 is affected by thrombi but classified as not having any. The reason can be clearly seen in \Cref{fig:M10}: the thrombus outside the lumen is very homogeneous, resulting in a very straight and regular lumen, with no bumps. 
By construction, our methodology is good at capturing information contained in inwards and outwards bumps, so the persistence diagrams of patient $A10$ are similar to those of a healthy patient. However, we do believe that segmenting also the thrombi from the CTA scans could solve this issue. We leave this further investigation to future works.

\section{Software and Packages}\label{sec:softw}

The segmentation pipeline makes use of \textit{vmtk 1.4.0} \cite{antiga2008image}, a collection of libraries and tools for 3D reconstruction and surface data analysis of blood vessels 
\cite{3_ANTIGA}.

The software used to visualize all the meshes and \acrshort{cta} scans shown in these pages is \textit{Paraview} \cite{ahrens200536} version \textit{5.11.0}.

Persistence diagrams and related distances are computed using the python library \textit{gudhi} \cite{maria2014gudhi} version 3.8.0, while for clusterization and visualization purposes the libraries \textit{scikit-learn} \cite{pedregosa2011scikit}, \textit{scipy} \cite{virtanen2020scipy}, \textit{seaborne} \cite{waskom2021seaborn} and \textit{matplotlib} \cite{hunter2007matplotlib} have been adopted.

\section{Discussion and conclusion}\label{sec:discussion}

We presented a complex pipeline to work with data obtained from medical imaging of abdominal aortas, exploiting the tools of topological data analysis. Each reconstructed mesh of an abdominal aorta is represented by an object called persistence diagram, displaying the 0-cycles and the 1-cycles characterizing a radial filtration function defined on the mesh. These diagrams are able to concisely summarize many shape-related features of the blood vessel, related to the irregularities of its lumen, which appear as a consequence of abdominal aortic aneurysms and some related features, like calcifications and intra-luminar thrombi. 

We argue that the proposed representation could lead to advances in the data analysis of AAAs. To support our claim we exhibit the results of several classification exercises, built upon a training set of persistent diagrams derived by 48 CTA scans, 24 of which picture a aneurysmal aorta. 

The take home message is that relying on a persistence diagram representation makes most of these classification exercises almost straightforward, due to the amount of information collected by these easy-to-handle mathematical objects. Moreover, because of the unsettling difference with more classical feature-gathering methods, we devoted a consistent part of this work to stress the interpretability of the rich information displayed by persistence diagrams.

Natural further developments of this work will tackle the forecasting and regression problems which have already captured the attention of the medical and statistical communities, like estimating growth rates of AAA, indication to surgery and rupture's risks. To that end, persistence diagrams present a number of mathematical advantages, briefly mentioned in the manuscript, allowing for more refined object oriented data analysis which will be the focus of our future works.  

\section{Acknowledgements}
Piercesare Secchi acknowledges the support by MUR, grant Dipartimento di Eccellenza 2023-2027.

Maurizio Domanin ackowledges the support of Irene Fulgheri for the help with Fondazione IRCCS Cà Granda Ospedale
Maggiore Policlinico's ethical committee.

\newpage

\appendix


\section{Average Neck Radius}\label{app:neck_radius}

Neck’s average radius has been estimated with the following steps:

\begin{enumerate}
    \item Linear interpolation of the centerline’s points.
    \item For a given parameter $s=0.5 mm$, empirically chosen, starting at the neck seed's height $s_0$ on the centerline, find the points $s_i$ at distance $s$ on the curvilinear abscissa.

    \item For each of the $s_i$ points find the planes $P_i$. orthogonal to the centerline in $s_i$. For each of the planes $P_i$ identified, a \textit{fitting ellipse} of the mesh points is built, following the Fitzgibbon’s approach \cite {4_ellipse1, 4_ellipse2}, and its semi-major and semi-minor axes $M_i$ and $m_i$ are computed. In some cases the slice can identify more regions; when this verifies, the ellipse is built using only the region of points that contains the point $s_i$.

    \item Starting from the initial seed $s_0$, a check is made for all the planes $P_j$ found. If the two proposition: 
    \begin{subequations}
        \begin{align}
        \frac{|M_i - M_{i+j}|}{M_i}<0.10  \quad \forall j=1,\cdots,k \quad \textit{ with k=10 (5 mm)}
        \end{align}
        \begin{align}
        M_i < 22.5 mm
        \end{align}

    \end{subequations}

    hold, $M_i$ and $m_i$ are collected. The first and all the following  $s_i$ that do not respect the proposition are discarded. k is chosen empirically and the conditions are made in order to control both the raw \acrshort{aaa}'s diameter and its relative growth.

   \item The result is the mean of the mean of $M_i$ and $m_i$ for each $s_i$ that is not discarded. In this way we obtain an accurate approximation of the neck's radius.
\end{enumerate}

\bibliography{references}


\end{document}